\documentclass[authoryear,3p]{elsarticle}
\makeatletter
\def\ps@pprintTitle{%
 \let\@oddhead\@empty
 \let\@evenhead\@empty
 \def\@oddfoot{\centerline{\thepage}}%
 \let\@evenfoot\@oddfoot}
\makeatother

\usepackage[utf8]{inputenc}
\usepackage{graphics}
\usepackage{graphicx}
\usepackage{amsmath,amssymb,esint}
\usepackage{lineno}
\usepackage{multirow}
\usepackage{subfig}

\usepackage{natbib}
\usepackage{eucal}
\usepackage{mathtools}
\usepackage{placeins}
\usepackage{enumerate}
\usepackage[colorlinks]{hyperref}

\usepackage[usenames,dvipsnames]{xcolor}

\usepackage[usenames,dvipsnames]{xcolor}

\journal{Journal of Sound and Vibration}

\begin{document}

\begin{frontmatter}

\title{Explicit Predictor-Corrector Method for Nonlinear Acoustic Waves Excited by a Moving Wave Emitting Boundary}

\author[1]{S{\"o}ren \textsc{Schenke} \corref{cor1}}
\author[1]{Fabian \textsc{Sewerin}}
\author[1]{Berend \textsc{van Wachem}}
\author[1]{Fabian \textsc{Denner}}

\address[1]{Chair of Mechanical Process Engineering, Otto-von-Guericke-Universit\"{a}t Magdeburg,\\ Universit\"atsplatz 2, 39106 Magdeburg, Germany}

\cortext[cor1]{Corresponding Author: Email: soeren.schenke@ovgu.de}

\begin{abstract}
We present an explicit finite difference time domain method to solve the lossless Westervelt equation for a moving wave emitting boundary in one dimension and in spherical symmetry. The approach is based on a coordinate transformation between a moving physical domain and a fixed computational domain. This allows to simulate the combined effects of wave profile distortion due to the constitutive nonlinearity of the medium and the nonlinear Doppler modulation of a pressure wave due to the acceleration of the wave emitting boundary. A predictor-corrector method is employed to enhance the numerical stability of the method in the presence of shocks and grid motion. It is demonstrated that the method can accurately predict the Doppler shift of nonlinear wave distortion and the amplitude modulation caused by an oscillating motion of the wave emitting boundary. The novelty of the presented methodology lies in its capability to reflect the Doppler shift of the rate of nonlinear wave profile distortion and shock attenuation for finite amplitude acoustic waves emitted from an accelerating boundary. 
\end{abstract}

\begin{keyword}
Nonlinear acoustics \sep Westervelt equation \sep moving boundary \sep Doppler effect \sep explicit finite difference \sep shock waves\\~\\
\textcopyright~2022. This manuscript version is made available under the CC-BY-NC-ND 4.0 license. \href{http://creativecommons.org/licenses/by-nc-nd/4.0/}{http://creativecommons.org/licenses/by-nc-nd/4.0/}
\end{keyword}

\end{frontmatter}



\section{Introduction}\label{sec:Introduction}

The physical theories describing the cumulative deformation, shock formation, and attenuation of finite amplitude acoustic waves in a quiescent fluid, where the acoustic wave is emitted by a stationary source, are well established \citep{Fay_1931, Westervelt_1963, Blackstock_1966}. In such a situation, the amplitude decay of the propagating wave is predominantly governed by the constitutive nonlinearity of the medium and the energy dissipation across the shock front. In the context of small amplitude wave propagation, recent developments \citep{Christov_2017, Gasperini_et_al_2021} provide us with a more complete picture of the nonlinear Doppler modulation of waves emitted from accelerating boundaries. Here, the constitutive nonlinearity of the medium is negligible, whereas the wave amplitude modulation induced by nonlinear Doppler effects plays an important role. In the present study, we present a modeling approach, extending the recent work by \citet{Gasperini_et_al_2021} to nonlinear waves, along with a numerical procedure, adopted from the work by \citet{Dey_and_Dey_1983} and \citet{Nascimento_et_al_2010}, that can reflect the combined effects of nonlinear wave profile distortion and nonlinear Doppler modulation.

The Westervelt equation proves to be useful for the modeling of nonlinear wave propagation \citep{Purrington_and_Norton_2012, Solovchuk_et_al_2013}, such as acoustic waves, if the formation of steep wave fronts is governed by cumulative nonlinear distortions during the wave propagation over several wavelengths \citep{Huijssen_phd_2008} due to the pressure-dependence of the particle velocity \citep{Hamilton_and_Blackstock_1988}. The Westervelt equation is given by \citep{Westervelt_1963}
\begin{equation}
\dfrac{\partial^2p}{\partial t^2}
-
\dfrac{\delta}{c_0^2} \dfrac{\partial^3p}{\partial t^3}
-
\dfrac{\beta}{\rho_0 c_0^2} \dfrac{\partial^2}{\partial t^2} \left(p^2\right) 
= 
c_0^2 \nabla^2 p,
\label{Westervelt_eqn}
\end{equation}
where $p$ is the pressure, $c_0$ the small signal sound speed, $\rho_0$ the density, $\delta$ the sound diffusivity \citep{Shevchenko_and_Kaltenbacher_2015}, and $\beta$ the nonlinearity coefficient as described in the work by \citet{Hamilton_and_Blackstock_1988}. The nonlinear behavior of waves radiated from fixed acoustic sources has been studied exhaustively. In comparison to that, relatively little is known about the nonlinear propagation of acoustic waves that are radiated from moving acoustic sources, especially when it comes to the combination of nonlinear wave propagation and accelerating wave emitting boundaries. The motion of a wave emitting boundary relative to a homogeneous and stationary background medium causes the well known Doppler shift in wavelength and frequency. This, in turn, is thought to have an effect on the rate of nonlinear wave steepening. In addition, the wave frequency is subject to a modulation beyond the linear Doppler shift when the source accelerates relative to the medium, which also causes a certain amplitude modulation \citep{Christov_2017, Gasperini_et_al_2021}. A detailed understanding of the mechanisms leading to the modulation of pressure waves is of particular importance for biomedical applications, where the generation of higher harmonics is associated with a higher attenuation of acoustic energy, which can alter the deposition of heat in biological tissue \citep{Muir_and_Carstensen_1980, Bailey_et_al_2003, Qiao_et_al_2016}. In order to investigate the modulation of pressure waves in a nonlinear medium in the presence of an accelerating source causing a nonlinear Doppler modulation, we present a numerical solution method for Eq. \eqref{Westervelt_eqn} that synthesizes the following two means:
\begin{enumerate}[i)]
\item The transformation of the Westervelt equation from a one-dimensional (1d) moving physical domain into a fixed computational domain, with extension to three-dimensional (3d) spherical symmetry. This provides a generic framework to solve the Westervelt equation for a moving wave emitting boundary.
\item The explicit finite difference solution of the transformed Westervelt equation using an anti-dispersive predictor-corrector method \citep{Nascimento_et_al_2010, Dey_and_Dey_1983} to increase the numerical stability of the solution and to avoid dispersive numerical noise.
\end{enumerate}
Very recently, \citet{Gasperini_et_al_2021} presented a formulation of the linear wave equation for moving wave scattering boundaries. Their approach is based on a coordinate transformation that conveys a mapping from a fixed computational domain, in which the equation is solved numerically, to a physical domain subject to accelerating boundary motion. In the present work, a similar coordinate transformation is applied to the Westervelt equation. In order to test the accuracy of the numerical model, the present study is mostly limited to test cases for which analytical reference solutions are available. In particular, we investigate: 
\begin{enumerate}[i)]
\item The linear Doppler modulation (constant speed of the wave emitting boundary) of the linear wave ($\beta=0$) in 1d.
\item The nonlinear Doppler modulation (accelerating wave emitting boundary) of the linear wave ($\beta=0$) in 1d.
\item The linear Doppler modulation (constant speed of the wave emitting boundary) of the nonlinear wave\footnote{For brevity, the waves formed in a nonlinear medium are termed ``nonlinear waves".} ($\beta>0$) in 1d.
\item The fully nonlinear wave propagation (accelerating wave emitting boundary and $\beta>0$) in 3d spherical symmetry.
\end{enumerate}
The governing equation is solved with an explicit FDTD (Finite Difference Time Domain) method based on central differences for the spatial derivatives, also referred to as FTCS (Forward Time Centered Space). Explicit time domain methods are popular because of their straightforward implementation and their computational efficiency. On the minus side, FDTD methods in general and FTCS schemes in particular are subject to very stringent stability criteria. Furthermore, FTCS schemes in their basic form are unconditionally unstable with respect to numerical errors in a von Neumann sense \citep{Dubey_2016}. The accumulation of phase dispersion errors is one of the main sources of numerical dispersion in FDTD methods for propagating waves \citep{Kyriakou_et_al_2015}. Dispersive numerical noise is further amplified by dispersive leading error terms \citep{Bouche_2003}. \citet{Hixon_1997} demonstrated that the onset of dispersive noise due to dispersive leading error terms is delayed with increasing order of accuracy of the approximation of the spatial terms. The accumulation of numerical dispersive noise can also be counteracted by increasing the spatial resolution, which, however, increases the computational cost significantly if the domain involves a large number of wavelengths on which the dispersive noise can develop \citep{Jaros_et_al_2014}. In the present study, the problem of dispersive noise is further aggravated by the formation of shocks, which involve high harmonics and naturally lead to Gibbs noise in spectral methods \citep{Jing_et_al_2011}. \citet{MacCormack_1982} presented a predictor-corrector method based on two subsequent opposite-sided finite difference approximations in space, which can dampen numerical noise generated by discontinuities in the medium \citep{Na_et_al_2016}. The Lax-Friedrich scheme is another frequently used dissipative scheme, even though it is pointed out by \citet{Shampine_2005} that the Lax-Friedrich scheme does not dampen the highest frequencies associated with the computational mesh, while the overall behavior of the scheme is strongly diffusive \citep{Breuss_2004}. As a second-order extension, the Lax-Wendroff scheme \citep{Lax_Wendroff_1960} is less dissipative. However, \citet{Murawska_et_al_2013} showed for the explicit finite difference simulation of a dam-break problem, that the Lax-Wendroff scheme, and also the original version of the aforementioned MacCormack method, are prone to nonphysical oscillations near steep wave fronts, which they could reduce considerably by complementing the MacCormack method by a total variation diminishing (TVD) scheme. \citet{Amara_et_al_2013} observed similar artificial oscillations in the explicit finite difference simulation of a water hammer problem and used the method of artificial viscosity to stabilize the solution in the vicinity of the shock front. In the present work, the feasibility of an anti-dispersive predictor-corrector method \citep{Nascimento_et_al_2010} based on the work of \citet{Dey_and_Dey_1983} is investigated, for which superior stability properties have been demonstrated in the context of explicit finite difference solutions \citep{Dey_1999}.

The background medium is assumed to be at rest, such that the motion of the wave emitting boundary does not induce any flow velocities in the surrounding medium. Furthermore, we neglect the sound diffusivity ($\delta=0$ in Eq. \eqref{Westervelt_eqn}), whence the resulting equation is referred to as the lossless Westervelt equation \citep{Meesala_et_al_2020}. The reason for neglecting the sound diffusivity is our focus on the representation of shocks. For an infinite wave train, excited by a sinusoidal signal and progressively deforming into an idealized saw-tooth shape, the Fay solution \citep{Fay_1931} provides an exact solution for the shock amplitude decay. As an essential part of the present work, it will be shown that the predictor-corrector method by \citet{Dey_and_Dey_1983} applied to the explicit FDTD solution of the lossless Westervelt equation can reproduce the physical shock amplitude decay in such a situation. However, the wave envelope described by the Fay solution is only exact if the wave amplitude decay is exclusively dictated by the dissipation of energy across the shock front \citep{Rudnick_1952}. In contrast, the attenuation due to the diffusion term in the Westervelt equation is frequency dependent and acts on the entire waveform, resulting in additional attenuation which is difficult to distinguish from the part of the amplitude decay caused by the shock. The present methodology can be extended to include the diffusion term in the Westervelt equation, and we anticipate that the additional physical diffusion will promote numerical stability in the presence of discontinuities, possibly permitting a reduction in the numerical diffusivity introduced by the predictor-corrector time integrator. The amplitude of a linear wave ($\beta=0$ in Eq. \eqref{Westervelt_eqn}) modulated by a linear Doppler shift must remain constant in a lossless medium. For the nonlinear amplitude modulation caused by small motion of the wave emitting boundary, the analytical solution by \citet{Christov_2017} can be employed to predict the waveform.

In Sec. \ref{losslesssec:The lossless Westervelt Equation for a Moving Wave Emitting Boundary}, we first derive the lossless Westervelt equation for a moving wave emitting boundary by applying a suitable coordinate transformation to the individual terms, which allows to solve the equation in a fixed computational domain. The explicit finite difference discretization of the transformed equation, which involves the aforementioned predictor-corrector steps to reduce numerical dispersion, is subject of Sec. \ref{sec:Explicit Finite Difference Discretization}. In Sec. \ref{sec:Excitation Signal at the Wave emitting boundary}, the excitation functions that are used in the numerical test-cases are specified, involving a brief discussion of their features. The fluid and wave parameters, the numerical settings, and the simulation results are presented in Sec. \ref{sec:Verification}. The results are discussed in Sec. \ref{sec:Discussion}, followed by the conclusion in Sec. \ref{sec:Conclusion}.

The numerical solution methodology we present yields an accurate and robust tool for investigating the behavior of nonlinear waves emitted from an accelerating boundary. In conjunction with the oscillatory motion of spherical wave emitting boundaries, this also constitutes an important step towards the accurate description of pressure waves incurred by cavitation bubbles with possible applications in the context of medical ultrasound therapy and/or diagnostic ultrasound, where a detailed understanding of the behavior of pressure waves emitted by oscillating gas bubbles is crucial \citep{Miller_et_al_2012, Hendee_and_Ritenour_2002}.

\section{The lossless Westervelt Equation for a Moving Wave Emitting Boundary}\label{losslesssec:The lossless Westervelt Equation for a Moving Wave Emitting Boundary}

Similar to the approach presented by \citet{Gasperini_et_al_2021}, we define both a moving physical domain $\Omega\left(t\right)=\left[X\left(t\right),L\right]$, with fixed right boundary $L$, time-dependent left boundary $X\left(t\right)$, and corresponding spatial coordinate $x$, and a fixed computational domain $\Theta=\left[0,1\right]$ with corresponding coordinate $\xi$. The pressure wave is radiated from the left boundary of $\Omega\left(t\right)$ at $x=X\left(t\right)$, where the excitation pressure $p_{\mathrm{ex}}\left(t\right)$ is prescribed. The transformation of the lossless Westervelt equation from $\left(x,t\right)$ to $\left(\xi,t\right)$ is achieved by invoking a time-dependent coordinate transformation on $\Omega\left(t\right)$, such that
\begin{equation}
x: \: \left[0,1\right] \rightarrow \left[X\left(t\right),L\right], \: \left(\xi,t\right) \mapsto x\left(\xi,t\right).
\label{coordTrans}
\end{equation}
With a slight abuse in notation, the symbol $x$ is used to denote both the coordinate transformation and a particular spatial location. First, the nonlinear term of the lossless Westervelt equation is linearized. Subsequently, the coordinate transformation from $x$ to $\xi$ is carried out. The resulting equations are simplified by assuming $x\left(\cdot, t\right)$ to vary linearly, and a special treatment of the mixed spatial/temporal derivative, resulting from the coordinate transformation, is introduced. Subsequently, the method is further extended to 3d spherical symmetry. In the final part of this section, the transformed lossless Westervelt equation is presented.

\subsection{Linearization}\label{sec:Linearization}

The nonlinear term in Eq. \eqref{Westervelt_eqn} can be rewritten as
\begin{equation}
\dfrac{\partial^2}{\partial t^2} \left(p^2\right)
=
2\left(\dfrac{\partial p}{\partial t}\right)^2 + 2p \dfrac{\partial^2p}{\partial t^2}.
\label{nonLinTerm}
\end{equation}
There are multiple options to handle this term numerically. \citet{Ramos_and_Nava_2012} mention that a quadratic expression involving two roots is obtained by direct discretization of the left-hand side of Eq. \eqref{nonLinTerm}. In their explicit FDTD approach, \citet{Doinikov_et_al_2014} treat the quadratic term and the pressure $p$ on the right-hand side of Eq. \eqref{nonLinTerm} as explicit source terms, meaning that the corresponding expressions are computed based on the previous time levels. A similar approach is followed by \citet{Haigh_et_al_2012}, who employ backward differences to avoid the presence of unknowns in the finite difference approximations of the time derivatives in Eq. \eqref{nonLinTerm}. In the present work, we opt for a Newton linearization of the nonlinear term, because the so obtained expression circumvents the issue of multiple roots while still allowing to move terms to the left-hand side of the discretized equation. The Newton linearization of the nonlinear term gives \citep{Solovchuk_et_al_2013}
\begin{equation}
\begin{array}{lll}
\dfrac{\partial^2}{\partial t^2} \left(p^2\right)
& \approx &
4\left(\dfrac{\partial p}{\partial t}\right)^o\dfrac{\partial p}{\partial t} 
+
 2p^o\dfrac{\partial^2 p}{\partial t^2}+ 2 p\left(\dfrac{\partial^2 p}{\partial t^2}\right)^o 
\\[10pt]
&&-
2\left(\left(\dfrac{\partial p}{\partial t}\right)^o\right)^2 
- 
2p^o\left(\dfrac{\partial^2 p}{\partial t^2}\right)^o,
\end{array}
\label{Newton_lin}
\end{equation}
where the superscript $o$ indicates the previous time level. The linearized form of the Westervelt equation then becomes
\begin{flalign}
& A_0p 
+ A_1 \dfrac{\partial p}{\partial t}
+ A_2 \dfrac{\partial^2 p}{\partial t^2}
=
A_{\mathrm{L}} \dfrac{\partial^2p}{\partial x^2} + N,
\label{Westervelt_eqn_lin_compact} \\
\intertext{where}
& A_0 = -\dfrac{2\beta}{\rho_0 c_0^2} \left(\dfrac{\partial^2p}{\partial t^2}\right)^o,
\label{A0} \\
& A_1 = -\dfrac{4\beta}{\rho_0 c_0^2} \left(\dfrac{\partial p}{\partial t}\right)^o,
\label{A1} \\
& A_2 = 1 - \dfrac{2\beta}{\rho_0 c_0^2} p^o,
\label{A2} \\
& A_{\mathrm{L}} = c_0^2,
\label{A4} \\
& N = - \dfrac{2\beta}{\rho_0 c_0^2} \left[\left(\left(\dfrac{\partial p}{\partial t}\right)^o\right)^2 + \left(\dfrac{\partial^2p}{\partial t^2}\right)^o p^o\right].
\label{Nterm}
\end{flalign}
The coefficients $A_0$ and $A_1$ in Eq. \eqref{Westervelt_eqn_lin_compact} are related to the nonlinearity, the coefficient $A_2$ has a nonlinear contribution and is equal to one if $\beta=0$. $A_{\mathrm{L}}$ is the coefficient of the Laplacian, and $N$ is the nonlinear contribution associated with the previous time level that results from the linearization in Eq. \eqref{Newton_lin}.

\subsection{Coordinate Transformation}\label{sec:Coordinate Transformation}

Based on the coordinate transformation given in Eq. \eqref{coordTrans} and following the methodology presented by \citet{Gasperini_et_al_2021}, the pressure $p$ in physical coordinates $\left(x, t\right)$ is equal to the pressure $\mathcal{P}$ in computational coordinates $\left(\xi, t\right)$, such that
\begin{equation}
p\left(x, t\right)
= 
\mathcal{P}\left(\xi\left(x, t\right), t\right).
\label{trans_def}
\end{equation}
Since the left-hand side of Eq. \eqref{trans_def} is a function of the free variable $x$, whereas the right-hand side of Eq. \eqref{trans_def} is a function of the dependent variable $\xi\left(x, t\right)$, differentiation with respect to $t$ gives
\begin{equation}
\dfrac{\partial p}{\partial t}
= 
\dfrac{\mathrm{d} \mathcal{P}}{\mathrm{d} t},
\label{dpdt_rule}
\end{equation}
where $\mathrm{d}/\mathrm{d}t$ is the total derivative operator, such that
\begin{equation}
\dfrac{\partial p}{\partial t}
= 
\dfrac{\partial \mathcal{P}}{\partial \xi}
\dfrac{\partial \xi}{\partial t}  
+ \dfrac{\partial \mathcal{P}}{\partial t}
\label{d1pdt1_x}
\end{equation}
and
\begin{equation}{}
\dfrac{\partial^2 p}{\partial t^2}
=
\dfrac{\partial^2 \mathcal{P}}{\partial \xi^2}  \left(\dfrac{\partial \xi}{\partial t}\right)^2
+
2\dfrac{\partial^2 \mathcal{P}}{\partial t \partial \xi} \dfrac{\partial \xi}{\partial t}
+
\dfrac{\partial \mathcal{P}}{\partial \xi} \dfrac{\mathrm{d}}{\mathrm{d} t} \left(\dfrac{\partial \xi}{\partial t} \right)
+
\dfrac{\partial^2 \mathcal{P}}{\partial t^2}.
\label{d2pdt2_x}
\end{equation}
Applying the chain rule to Eq. \eqref{trans_def} gives
\begin{equation}
\dfrac{\partial p}{\partial x} = \dfrac{\partial \xi}{\partial x} \dfrac{\partial \mathcal{P}}{\partial \xi}
\label{gradtrans}
\end{equation}
for the pressure gradient, where the term $\partial \xi/\partial x$ can be interpreted as the Jacobian of the coordinate transformation \citep{Bernard_1992, Liseikin_2017}. For the Laplacian, we get
\begin{equation}
\dfrac{\partial^2 p}{\partial x^2} 
=
\left(\dfrac{\partial \xi}{\partial x}\right)^2 \dfrac{\partial^2 \mathcal{P}}{\partial \xi^2}
+
\dfrac{\partial \mathcal{P}}{\partial \xi} \dfrac{\partial^2 \xi}{\partial x^2}.
\label{laplacetrans}
\end{equation}
Eqs. \eqref{d1pdt1_x} to \eqref{laplacetrans} are presented in the work by \citet{Gasperini_et_al_2021}.

\subsection{Piece-wise Linearity and Uniform Grids}\label{sec:Piece-wise Linearity and Uniform Grids}

In the present article, we consider spatially uniform grids in both $x$ and $\xi$, meaning that $\Delta \xi = \mathrm{const}$ and $\Delta x = \mathrm{const}$. For simplicity, the coordinate transformation in Eq. \eqref{coordTrans} is defined to vary linearly in $\xi$, such that
\begin{equation}
\xi\left(x,t\right) 
= 
\dfrac{x-X\left(t\right)}{L-X\left(t\right)}
\hspace{0.3cm}
\Leftrightarrow
\hspace{0.3cm}
x\left(\xi,t\right) = X\left(t\right) + \xi \left(L-X\left(t\right)\right).
\label{xifun}
\end{equation}
For the spatial derivatives given by Eqs. \eqref{gradtrans} and \eqref{laplacetrans}, this yields
\begin{equation}
\dfrac{\partial p}{\partial x} =
\dfrac{1}{L-X\left(t\right)} \dfrac{\partial \mathcal{P}}{\partial \xi},
\label{gradtrans2}
\end{equation}
and, as the second term on the right-hand side of Eq. \eqref{laplacetrans} vanishes due to the assumption that $\xi\left(x\right)$ is piece-wise linear in $x$,
\begin{equation}
\dfrac{\partial^2 p}{\partial x^2} =
\dfrac{1}{\left(L-X\left(t\right)\right)^2} \dfrac{\partial^2 \mathcal{P}}{\partial \xi^2}.
\label{laplacetrans2}
\end{equation}
The term $1/\left(L-X\left(t\right)\right)$ is associated with the Jacobian. It is further convenient to define
\begin{equation}
q = \dfrac{\partial \xi}{\partial t},
\label{qdef}
\end{equation}
where the quantity $q$ can be interpreted as the perceived velocity of a fixed position in the computational domain viewed from a (moving) location in physical space. Further assuming that the grid deformation is piece-wise linear in time as well, we can drop the higher-order time derivative $\mathrm{d}q/\mathrm{d}t$ in Eq. \eqref{d2pdt2_x}, such that Eqs. \eqref{d1pdt1_x} and \eqref{d2pdt2_x} simplify to
\begin{equation}
\dfrac{\partial p}{\partial t}
= 
q\dfrac{\partial \mathcal{P}}{\partial \xi}
+ \dfrac{\partial \mathcal{P}}{\partial t},
\label{d1pdt1_pwlin}
\end{equation}
\begin{equation}
\dfrac{\partial^2 p}{\partial t^2}
= 
q^2\dfrac{\partial^2 \mathcal{P}}{\partial \xi^2}
+
2q\dfrac{\partial^2 \mathcal{P}}{\partial t \partial \xi}
+
\dfrac{\partial^2 \mathcal{P}}{\partial t^2}.
\label{d2pdt2_pwlin}
\end{equation}
Differentiating $\xi\left(x\left(\xi,t\right),t\right)=\xi$ with respect to time gives the identity 
\begin{equation}
\dfrac{\partial \xi}{\partial t}
+
\dfrac{\partial \xi}{\partial x}\dfrac{\partial x}{\partial t}
= 0.
\label{xiIdent}
\end{equation}
From Eqs. \eqref{xiIdent} and \eqref{qdef}, and the equivalent expressions in Eq. \eqref{xifun}, it follows that
\begin{equation}
q\left(\xi,t\right)
=
- \dfrac{1-\xi}{L-X\left(t\right)} \dfrac{\partial X}{\partial t}.
\label{xitrans3}
\end{equation}
In order to enhance the stability of the explicit finite difference solution, it is advantageous to replace the mixed temporal/spatial derivative in Eq. \eqref{d2pdt2_pwlin} by an equivalent expression that can partially be moved to the left-hand side of the discretized equation. Such an equivalent expression can be deduced by applying the mixed derivative operator to the product of $q\left(\xi, t\right)$ and $\mathcal{P}\left(\xi, t\right)$, such that
\begin{equation}
\dfrac{\partial^2}{\partial t \partial \xi} \left[q\left(\xi, t\right)\mathcal{P}\left(\xi, t\right)\right]
=
\dfrac{\partial^2 q}{\partial \xi^2}q  \mathcal{P}
+
\dfrac{\partial q}{\partial \xi} \dfrac{\partial  \mathcal{P}}{\partial t}
+
\dfrac{\partial q}{\partial t} \dfrac{\partial  \mathcal{P}}{\partial \xi}
+
q \dfrac{\partial^2 \mathcal{P}}{\partial t \partial \xi}.
\label{dtdxi_qp}
\end{equation}
Eq. \eqref{dtdxi_qp} can be rearranged for the last term on the right-hand side, which is the mixed derivative term in Eq. \eqref{d2pdt2_pwlin}. It is noted that no additional coordinate transformation is involved in Eq. \eqref{dtdxi_qp}. The term $\partial q/\partial \xi$ is the divergence of $q$ with respect to $\xi$, which is constant in $\xi$ if $x\left(\xi, t\right)$ is a linear function in $\xi$. Hence, $\partial^2 q/\partial \xi^2 = 0$. Again assuming piece-wise linearity in time, the mixed term in Eq. \eqref{d2pdt2_pwlin} can be replaced by
\begin{equation}
q \dfrac{\partial^2 \mathcal{P}}{\partial \xi \partial t}
=
\dfrac{\partial^2}{\partial t \partial \xi}\left(q\mathcal{P}\right)
-
\dfrac{\partial q}{\partial \xi} \dfrac{\partial  \mathcal{P}}{\partial t}.
\label{dtdx_equiv}
\end{equation}
Eq. \eqref{dtdx_equiv} still includes a mixed derivative, but also an additional unmixed contribution, which can partially be moved to the left-hand side of the discretized equation. Substituting Eq. \eqref{dtdx_equiv} into Eq. \eqref{d2pdt2_pwlin}, the second partial pressure derivative takes the final form
\begin{equation}
\dfrac{\partial p^2}{\partial t^2}
= 
q^2\dfrac{\partial^2 \mathcal{P}}{\partial \xi^2}
+
2\dfrac{\partial^2}{\partial t \partial \xi}\left(q\mathcal{P}\right)
-
2\dfrac{\partial q}{\partial \xi} \dfrac{\partial  \mathcal{P}}{\partial t}
+
\dfrac{\partial^2 \mathcal{P}}{\partial t^2},
\label{d2pdt2_pwlin_final}
\end{equation}
where the third term and the fourth term on the right-hand side will be moved to the left hand side of the discretized equation. It is noted that due to the time derivative of $q\mathcal{P}$, Eq. \eqref{d2pdt2_pwlin_final} is not strictly piece-wise linear in time, but involves a higher order contribution.

\subsection{Extension to Spherical Symmetry}\label{sec:Extension to Spherical Symmetry}

The equations are extended to 3d spherical coordinates under the assumption of spherical symmetry, so that the algebraic form of the equations remains 1d. The coordinate $x$ is replaced by the radial coordinate $r$, and the position of the wave emitting boundary, previously denoted by $X\left(t\right)$, is now denoted by $R\left(t\right)$. In 3d spherical symmetry, the Laplacian in the original Westervelt Eq. \eqref{Westervelt_eqn} becomes \citep{Whitham_1999}
\begin{equation}
\nabla^2 p = \dfrac{2}{r} \dfrac{\partial p}{\partial r} + \dfrac{\partial^2p}{\partial r^2}.
\label{sphericalLaplacian}
\end{equation}
With the coordinate transformation
\begin{equation}
r\left(\xi\right) = R\left(t\right) + \xi \left(L-R\left(t\right)\right),
\label{xifun_rad}
\end{equation}
we obtain
\begin{equation}
\dfrac{\partial p}{\partial r} =
\dfrac{1}{L-R\left(t\right)} \dfrac{\partial \mathcal{P}}{\partial \xi}.
\label{gradtrans3}
\end{equation}
Since Eq. \eqref{xifun_rad} is identical to Eq. \eqref{xifun} except for the change from $x$ to $r$, the transformations of the time derivatives of $p\left(r,t\right)$ remain unaffected.

\subsection{Transformed Westervelt Equation}\label{sec:Transformed Westervelt Equation}

With the transformations of the time and space derivatives of $p$ given by Eqs. \eqref{laplacetrans2}, \eqref{d1pdt1_pwlin}, and \eqref{d2pdt2_pwlin_final}, the linearized Westervelt equation and its coefficients and coupling terms follow as
\begin{flalign}
& \mathcal{A}_0 \mathcal{P}
+ \mathcal{A}_1  \dfrac{\partial \mathcal{P}}{\partial t}
+ \mathcal{A}_2  \dfrac{\partial^2 \mathcal{P}}{\partial t^2}
=
\mathcal{A}_{\mathrm{L}}\dfrac{\partial^2\mathcal{P}}{\partial \xi^2} 
+ \left(\mathcal{A}_{\mathrm{G}} + \mathcal{A}_{\mathrm{G,r}}\right)\dfrac{\partial\mathcal{P}}{\partial \xi}
+ \mathcal{N} + \mathcal{K},
\label{LinEqn}
\\
\intertext{where}
& \mathcal{A}_0 = -\dfrac{2\beta}{\rho_0 c_0^2}
\left(\dfrac{\mathrm{d}^2\mathcal{P}}{\mathrm{d}t^2}\right)^o,
\label{A0_defGrid}
\\
& \mathcal{A}_1 = -\dfrac{4\beta}{\rho_0 c_0^2} \left(\dfrac{\mathrm{d}\mathcal{P}}{\mathrm{d}t}\right)^o - 2\left(1 - \dfrac{2\beta}{\rho_0 c_0^2} \mathcal{P}^o\right)\dfrac{\partial q}{\partial \xi},
\label{A1_defGrid} \\
& \mathcal{A}_2 = 1 - \dfrac{2\beta}{\rho_0 c_0^2} \mathcal{P}^o,
\label{A2_defGrid} \\
& \mathcal{A}_{\mathrm{L}} = \dfrac{c_0^2}{\left(L-X\left(t\right)\right)^2} - \left(1 - \dfrac{2\beta}{\rho_0 c_0^2} \mathcal{P}^o\right)q^2,
\label{A4_defGrid} \\
& \mathcal{A}_{\mathrm{G}} = \dfrac{4\beta}{\rho_0 c_0^2}
\left(\dfrac{\mathrm{d}\mathcal{P}}{\mathrm{d}t}\right)^o q,
\label{coeffAG} \\
& \mathcal{A}_{\mathrm{G},r} = \dfrac{2c_0^2}{\left(X\left(t\right) + \xi \left(L-X\left(t\right)\right)\right)\left(L-X\left(t\right)\right)},
\label{coeffAr} \\
& \mathcal{N} = - \dfrac{2\beta}{\rho_0 c_0^2} \left[\left(\left(\dfrac{\mathrm{d}\mathcal{P}}{\mathrm{d}t}\right)^o\right)^2 + \left(\dfrac{\mathrm{d}^2\mathcal{P}}{\mathrm{d}t^2}\right)^o \mathcal{P}^o\right],
\label{Nterm_defGrid} \\
&\mathcal{K} =
2\left(1 - \dfrac{2\beta}{\rho_0 c_0^2}\right)\dfrac{\partial^2}{\partial t \partial \xi} \left(q \mathcal{P}\right),
\label{kinCoupling} \\
&\left(\dfrac{\mathrm{d}\mathcal{P}}{\mathrm{d}t}\right)^o =
\left(q\dfrac{\partial \mathcal{P}}{\partial \xi}
+ \dfrac{\partial \mathcal{P}}{\partial t}\right)^o,
\label{dPdto} \\
&\left(\dfrac{\mathrm{d}^2\mathcal{P}}{\mathrm{d}t^2}\right)^o =
\left(q^2\dfrac{\partial^2 \mathcal{P}}{\partial \xi^2}
+
2q\dfrac{\partial^2 \mathcal{P}}{\partial t \partial \xi}
+
\dfrac{\partial^2 \mathcal{P}}{\partial t^2}\right)^o.
\label{d2Pdt2o}
\end{flalign}
The coefficient $\mathcal{A}_2$ is identical to the coefficient $A_2$ (Eq. \eqref{A2}) of the linearized Westervelt Eq. \eqref{Westervelt_eqn_lin_compact}, except for the change from $p$ to $\mathcal{P}$. The coefficients $\mathcal{A}_0$, $\mathcal{A}_1$, and $\mathcal{N}$ correspond to the coefficients $A_0$, $A_1$, and $N$ in Eq. \eqref{Westervelt_eqn_lin_compact}. However, the corresponding time derivatives are subject to the coordinate transformation from $x$ to $\xi$, leading to Eqs. \eqref{d1pdt1_pwlin} and \eqref{d2pdt2_pwlin}. Eqs. \eqref{dPdto} and \eqref{d2Pdt2o} are used to approximate the time derivatives at the previous time levels as given by Eqs. \eqref{dPdto} and \eqref{d2Pdt2o}. The Laplacian coefficient $\mathcal{A}_{\mathrm{L}}$ given by Eq. \eqref{A4_defGrid} has two contributions. The first contribution corresponds to the original coefficient $A_{\mathrm{L}}$ in Eq. \eqref{A4}, however scaled by the square of the Jacobian of the coordinate transformation. The second contribution results from the coordinate transformation of the partial pressure derivatives, which also yields the additional gradient term given by Eq. \eqref{coeffAG} with the corresponding coefficient $\mathcal{A}_{\mathrm{G}}$, as well as the additional kinematic coupling term $\mathcal{K}$ given by Eq. \eqref{kinCoupling}. The coefficient $\mathcal{A}_{\mathrm{G},r}$ given by Eq. \eqref{coeffAr} is only required in spherical coordinates and is equal to zero in the Cartesian case. For convenience, $X\left(t\right)$ is formally replaced by $R\left(t\right)$ in the case of 3d spherical symmetry (see Sec. \ref{sec:Extension to Spherical Symmetry}).

\section{Explicit Finite Difference Discretization}\label{sec:Explicit Finite Difference Discretization}

In the following, the discretization procedure and the initial and boundary conditions are presented. The explicit FDTD method is complemented by an anti-dispersive predictor-corrector step \citep{Dey_and_Dey_1983, Nascimento_et_al_2010} to counteract the build-up of dispersive numerical noise due to rapid grid motion and/or the presence of shocks. The corrected solution is based on a re-evaluation of the Laplacian, and the final solution is then given by a weighted sum of the predicted and the corrected solution.

\subsection{Finite Difference Approximation}\label{sec:Finite Difference Approximation}

Concerning the finite difference approximations of the spatial derivatives, $4^{\mathrm{th}}$-order accurate central differences are commonly used to solve the Westervelt equation with the explicit FDTD method \citep{Hallaj_and_Cleveland_1999, Norton_and_Purrington_2009, Doinikov_et_al_2014, Karamalis_et_al_2010}. In the present work, a $6^{\mathrm{th}}$-order accurate central difference scheme is used, since the higher order can further delay the onset of dispersive numerical noise as shown by \citet{Hixon_1997}. Using the central difference coefficients found in the work by \citet{Fornberg_1988}, the $6^{\mathrm{th}}$-order accurate central finite difference approximations of the first and the second spatial derivatives of the pressure $\mathcal{P}$ with respect to the computational coordinate $\xi$ are given by
\begin{equation}
\left(\dfrac{\partial \mathcal{P}}{\partial \xi}\right)_i^j
\approx 
\dfrac{1}{\Delta \xi}\left(
-\dfrac{1}{60} \mathcal{P}_{i-3}^j
+\dfrac{3}{20} \mathcal{P}_{i-2}^j
-\dfrac{3}{4} \mathcal{P}_{i-1}^j
+\dfrac{3}{4} \mathcal{P}_{i+1}^j
-\dfrac{3}{20} \mathcal{P}_{i+2}^j
+\dfrac{1}{60} \mathcal{P}_{i+3}^j
\right) + \mathcal{O}\left(\Delta \xi^6\right),
\label{FD_ddx1}
\end{equation}
\begin{equation}
\left(\dfrac{\partial^2 \mathcal{P}}{\partial \xi^2}\right)_i^j
\approx 
\dfrac{1}{\Delta \xi^2}\left(
\dfrac{1}{90} \mathcal{P}_{i-3}^j
-\dfrac{3}{20} \mathcal{P}_{i-2}^j
+\dfrac{3}{2} \mathcal{P}_{i-1}^j
-\dfrac{49}{18} \mathcal{P}_{i}^j
+\dfrac{3}{2} \mathcal{P}_{i+1}^j
-\dfrac{3}{20} \mathcal{P}_{i+2}^j
+\dfrac{1}{90} \mathcal{P}_{i+3}^j
\right) + \mathcal{O}\left(\Delta \xi^6\right),
\label{FD_ddx2}
\end{equation}
where $i$ and $j$ indicate the discrete location and time instant, respectively. For the time derivatives, we employ the approximations
\begin{equation}
\left(\dfrac{\partial \mathcal{P}}{\partial t}\right)^{j}_i 
=
\dfrac{1}{\Delta t}
\left(
 \mathcal{P}^{j+1}_i
-\mathcal{P}^{j}_i
\right) + \mathcal{O}\left(\Delta t^1\right),
\label{fddt1}
\end{equation}
\begin{equation}
\left(\dfrac{\partial^2 \mathcal{P}}{\partial t^2}\right)^{j}_i 
=
\dfrac{1}{\Delta t^2}
\left(
 \mathcal{P}^{j+1}_i
-2 \mathcal{P}^{j}_i
+\mathcal{P}^{j-1}_i
\right) + \mathcal{O}\left(\Delta t^2\right).
\label{fddt2}
\end{equation}
The approximation of the second time derivative given by Eq. \eqref{fddt2} is associated with a central finite difference around the time instant $j$ and is therefore $2^{\mathrm{nd}}$-order accurate \citep{Haigh_et_al_2012}, whereas the approximation of the first time derivative given by Eq. \eqref{fddt1} is $1^{\mathrm{st}}$-order accurate. The choice of the $1^{\mathrm{st}}$-order approximation follows the predictor-corrector method (see Sec. \ref{sec:Predictor-Corrector Method}) developed by \citet{Dey_and_Dey_1983}, where the predictor step is based on a forward Euler step.

\subsection{Predictor-Corrector Method}\label{sec:Predictor-Corrector Method}

In more compact notation, where the spatial index $i$ is omitted for brevity, the finite difference approximations of the time derivatives given by Eqs. \eqref{fddt1} and \eqref{fddt2} can be written as
\begin{equation}
\left(\dfrac{\partial^k \mathcal{P}}{\partial t^k}\right)^{j}
=
\dfrac{a_0^{\left(k\right)}}{\Delta t^k} \mathcal{P}^{j+1} + \dfrac{b^{\left(k,j\right)}}{\Delta t^k},
\hspace{0.3cm} \textnormal{where} \hspace{0.3cm}
b^{\left(k,j\right)} = \sum_{n=1}^{N_{\mathrm{coeff}}-1} a_n^{\left(k\right)}\mathcal{P}^{j+1-n},
\label{fddt_compact}
\end{equation}
where $k$ indicates the derivative order, $a_n^{\left(k\right)}$ the corresponding finite difference coefficients, and $N_{\mathrm{coeff}}$ the number of finite difference coefficients for the corresponding stencil. With Eq. \eqref{fddt_compact}, Eq. \eqref{LinEqn} takes the semi-discrete form
\begin{equation}
\begin{array}{lll}
\left(
\mathcal{A}_0^j 
+
\displaystyle \sum_{k=1}^2 \mathcal{A}_k^j \dfrac{a_0^{\left(k\right)}}{\Delta t^k}
\right) \mathcal{P}^{j+1}
&=&
\mathcal{A}_{\mathrm{L}}^j\left(\dfrac{\partial^2\mathcal{P}}{\partial \xi^2}\right)^j 
+
\left(\mathcal{A}_{\mathrm{G}}^j + \mathcal{A}_{\mathrm{G},r}^j\right)\left(\dfrac{\partial\mathcal{P}}{\partial \xi}\right)^j 
\\[10pt]
&&+
\mathcal{N}^j
+
\mathcal{K}^j
- 
\displaystyle \sum_{k=1}^2 \mathcal{A}_k^j \dfrac{b^{\left(k,j\right)}}{\Delta t^k}.
\end{array}
\label{LinEqn_discr}
\end{equation}
Eq. \eqref{LinEqn_discr} can be solved for the new pressure $\mathcal{P}^{j+1}$ explicitly. With the substitutions
\begin{equation}
\mathfrak{A}^j =
\left(
\mathcal{A}_0^j 
+
\sum_{k=1}^2 \mathcal{A}_k^j \dfrac{a_0^{\left(k\right)}}{\Delta t^k}
\right),
\label{defAA}
\end{equation}
\begin{equation}
\mathfrak{B}^j =
\mathcal{A}_{\mathrm{G}}^j\left(\dfrac{\partial\mathcal{P}}{\partial \xi}\right)^j
+
\mathcal{N}^j
+
\mathcal{K}^j
-
\sum_{k=1}^2 \mathcal{A}_k^j \dfrac{b^{\left(k,j\right)}}{\Delta t^k},
\label{defBB}
\end{equation}
Eq. \eqref{LinEqn_discr} is further simplified to
\begin{equation}
\widetilde{\mathcal{P}}
=
\dfrac{1}{\mathfrak{A}^j}
\left[
\mathcal{A}_{\mathrm{L}}^j\left(\dfrac{\partial^2\mathcal{P}}{\partial \xi^2}\right)^j
+
\mathcal{A}_{\mathrm{G},r}^j\left(\dfrac{\partial\mathcal{P}}{\partial \xi}\right)^j 
+
\mathfrak{B}^j
\right].
\label{LinEqn_discr2}
\end{equation}
Following the approach by \citet{Dey_and_Dey_1983}, the solution $\widetilde{\mathcal{P}}$ obtained from Eq. \eqref{LinEqn_discr2} is considered as an interim/predicted solution, which is then corrected by
\begin{equation}
\mathcal{P}^{j+1}
=
\left(1-\gamma\right)\widetilde{\mathcal{P}}
+
\dfrac{\gamma}{\mathfrak{A}^j}
\left[
\mathcal{A}_{\mathrm{L}}^j\dfrac{\partial^2\widetilde{\mathcal{P}}}{\partial \xi^2}
+
\mathcal{A}_{\mathrm{G},r}^j \dfrac{\partial\widetilde{\mathcal{P}}}{\partial \xi}
+
\mathfrak{B}^j
\right]
\hspace{0.2cm}
\textnormal{for}
\hspace{0.2cm}
\gamma \in \left[0,1\right],
\label{LinEqn_discr_corr}
\end{equation}
where only the Laplacian has been updated after the predictor step, similar to the approach by \citet{Nascimento_et_al_2010} who applied the predictor-corrector method to the linear wave equation. In Eq. \eqref{LinEqn_discr_corr}, the predicted and the corrected solutions are weighted by the relaxation factor $\gamma$. For $\gamma=0$, the scheme corresponds to the standard FTCS scheme. By substituting the predictor and the corrector equations into each other and by recasting the so obtained equation back into the associated continuous wave equation, \citet{Nascimento_et_al_2010} showed that the predictor-corrector method effectively introduces an additional damping term. Performing analogous steps as in the work by \citet{Nascimento_et_al_2010}, i.e., substituting Eq. \eqref{LinEqn_discr2} into Eq. \eqref{LinEqn_discr_corr}, it follows, after some manipulations, that the so obtained equation corresponds to the discretized form of the new wave equation
\begin{equation}
\begin{array}{ll}
\mathcal{A}_0 \mathcal{P}
+ \mathcal{A}_1  \dfrac{\partial \mathcal{P}}{\partial t}
+ \mathcal{A}_2  \dfrac{\partial^2 \mathcal{P}}{\partial t^2}
&=
\mathcal{A}_{\mathrm{L}}\dfrac{\partial^2\mathcal{P}}{\partial \xi^2} 
+ \left(\mathcal{A}_{\mathrm{G}} + \mathcal{A}_{\mathrm{G,r}}\right)\dfrac{\partial\mathcal{P}}{\partial \xi}
+ \mathcal{N} + \mathcal{K} \\[10pt]
& + \gamma \Delta t\left[\mathcal{A}_{\mathrm{L}} \dfrac{\partial}{\partial t}\left( \dfrac{\partial^2\mathcal{P}}{\partial \xi^2}\right)
+\mathcal{A}_{\mathrm{G,r}} \dfrac{\partial}{\partial t}\left( \dfrac{\partial\mathcal{P}}{\partial \xi}\right)\right] + \gamma \mathcal{H}\left(\Delta t^k\right).
\end{array}
\label{LinEqn_antiDispersion}
\end{equation}
For $\gamma=0$, the transformed Westervelt equation given by Eq. \eqref{LinEqn} is recovered. The term in square brackets in Eq. \eqref{LinEqn_antiDispersion} is associated with the partial time derivative of the (spherical) Laplacian, which has a damping effect on the acoustic wave \citep{Shevchenko_and_Kaltenbacher_2015}. The magnitude of the time derivative of the (spherical) Laplacian increases with decreasing wavelength, so that the damping rate is biased towards higher wave frequencies, which in turn suppresses dispersion and causes an attenuation of higher harmonics, particularly, as near-discontinuities form (see Fig. \ref{fig:longSimu_dxStudy} below). Since the damping term is proportional to the corrector weight $\gamma$ and the time step size $\Delta t$, both its magnitude and the amount of induced artificial diffusion decrease with decreasing $\gamma$ or $\Delta t$.
As $\Delta t$ decreases, the stabilizing effect on the numerical solution is preserved because the decrease of $\Delta t$ shifts the onset of dispersive numerical noise to higher wave frequencies. Concomitantly, the spatial resolution ought to be adjusted in such a way that the higher wave harmonics can still be accurately resolved. This implies, in turn, that the Courant-Friedrichs-Lewy ($\mathrm{CFL}=c_0\Delta t/\Delta x$) number cannot be chosen arbitrarily small without impairing the stabilizing effect of the predictor-corrector method. The term $\mathcal{H}\left(\Delta t^k\right)$ in Eq. \eqref{LinEqn_antiDispersion} involves additional terms resulting from the substitution of Eq. \eqref{LinEqn_discr2} into Eq. \eqref{LinEqn_discr_corr} and essentially depends on higher orders of $\Delta t$. In the derivation by \citet{Nascimento_et_al_2010} for the linear wave equation, $\mathcal{H}$ involves only one single term. In the present work, the number of terms in $\mathcal{H}$ increases significantly due to the presence of the nonlinear term and the coupling terms resulting from the coordinate transformation on physical space. Without the need to specify $\mathcal{H}$, it is noted that $\mathcal{H}$ is implicitly taken into account in our approach by performing the predictor-corrector steps as given by Eqs. \eqref{LinEqn_discr2} and \eqref{LinEqn_discr_corr}, corresponding to the original predictor-corrector method by \cite{Dey_and_Dey_1983}. \citet{Nascimento_et_al_2010}, by contrast, omit $\mathcal{H}$ and only keep the additional damping term. Motivated by the objective to derive an anti-dispersion wave equation, \citet{Liu_et_al_2009} introduced a very similar damping term, however without relating it to the predictor-corrector method by \citet{Dey_and_Dey_1983}.

\subsection{Initial and Boundary Conditions}\label{sec:Initial and Boundary Conditions}

\begin{figure}[htb]
\begin{center}
{\includegraphics[width=\linewidth]{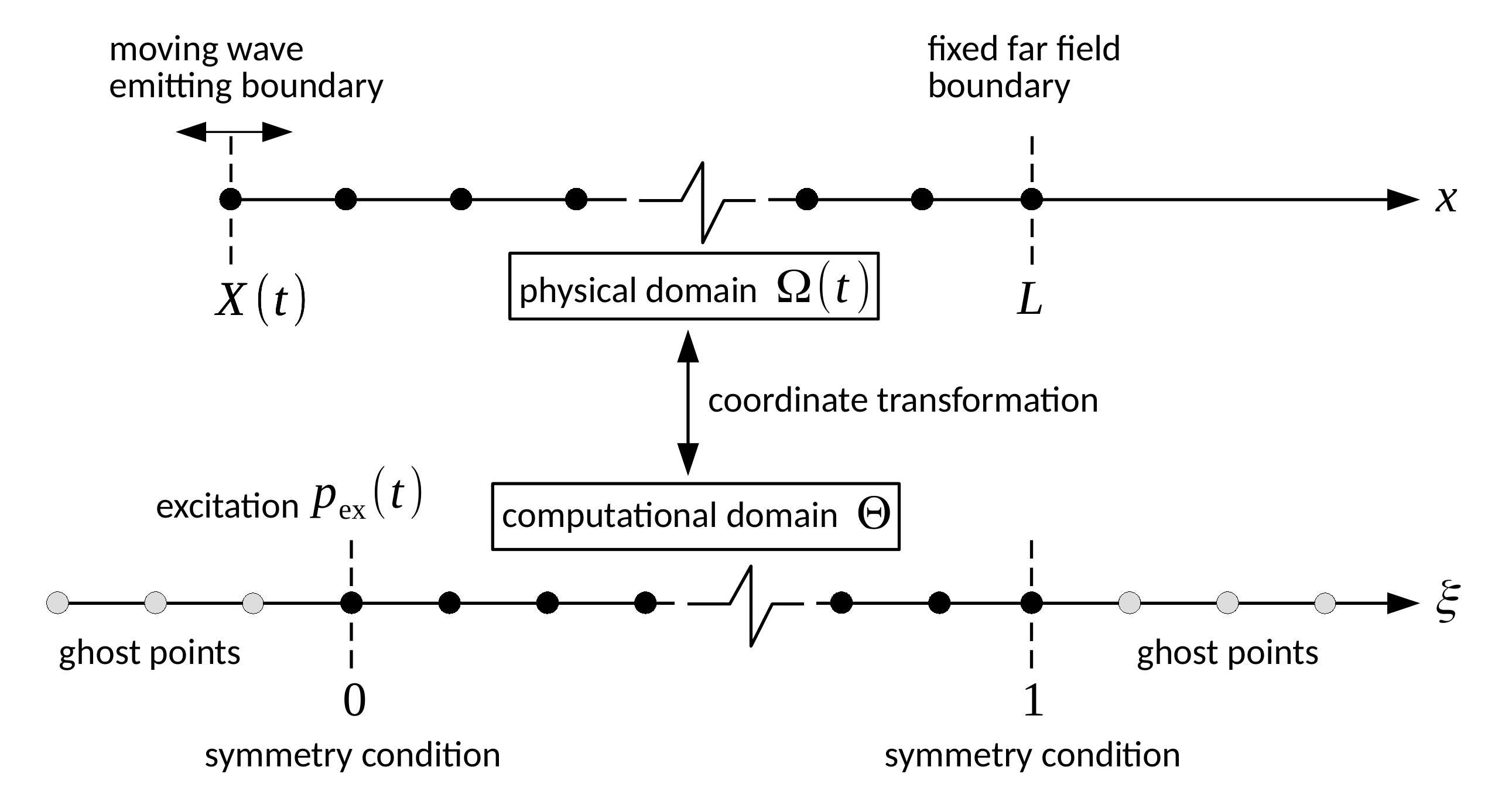}}
\caption{Sketch of the physical domain, with the moving wave emitting boundary, and the fixed computational domain, where the pressure wave is excited at the wave emitting boundary and where the ghost points are needed to establish the symmetry condition so that the central finite difference approximations given by Eqs. \eqref{FD_ddx1} and \eqref{FD_ddx2} can be applied near the domain boundaries. The physical domain size is chosen such that the acoustic waves do not reach the wave reflecting far field boundary.}
\label{fig:domainNew}
\end{center}
\end{figure}

The physical domain and the computational domain, related to each other by the time-dependent coordinate transformation given by Eq. \eqref{coordTrans}, are sketched in Fig. \ref{fig:domainNew}. In spherical symmetry, the Cartesian coordinate $x$ and the wave emitting boundary position $X\left(t\right)$ are formally replaced by the radial coordinate $r$ and the sphere radius $R\left(t\right)$ (see Sec. \ref{sec:Extension to Spherical Symmetry}). At the wave emitting boundary, a symmetry condition is imposed for all cases. The wave emitting boundary is treated as a grid point, at which the excitation pressure $p_{\mathrm{ex}}\left(t\right)$ is imposed as is further specified in Sec. \ref{sec:Excitation Signal at the Wave emitting boundary} below. In order to establish the symmetry condition, the three neighbouring grid points of the wave emitting boundary and the corresponding discrete pressure values are mirrored at the wave emitting boundary, so that a ghost wave is radiated in the opposite direction of the wave propagating into the computational domain. With this particular symmetry condition, the central differences of the spatial derivatives in the immediate proximity of the wave emitting boundary can be calculated with the same stencil and coefficients (see Eqs. \eqref{FD_ddx1} and \eqref{FD_ddx2}) as in the domain interior. The domain size and the simulation time are always chosen in such a way that the wave cannot reach the far field boundary, so that practically, the choice of the far field boundary condition is immaterial in the present study. This constraint on the domain size is necessary because we currently use the same symmetry condition at the far field boundary as for the wave emitting boundary, which would result in wave reflections as soon as the wave reaches the far field boundary. At the initial time, the acoustic pressure field is zero. A Gauss envelope as described in Sec. \ref{sec:Excitation Signal at the Wave emitting boundary} is applied to the excitation signal in order to circumvent start-up discontinuities.

\section{Excitation Signal at the Wave emitting boundary}\label{sec:Excitation Signal at the Wave emitting boundary}

In the present study, the pressure excitation takes place at the wave emitting boundary ($x=X\left(t\right)$ in 1d or $r=R\left(t\right)$ in 3d spherical symmetry), which is either at rest, moving at constant speed, or accelerating in an oscillatory fashion. The excitation pressure signal is represented by a sinusoidal function specified in Sec. \ref{sec:Pressure Excitation}. Therefore, either the pressure $p$, its gradient, or its Laplacian have a discontinuity at the initial time instant. As pointed out by \citet{Karamalis_et_al_2010}, this can lead to spurious disturbances in the numerical solution. As a countermeasure, the excitation signal is convoluted with a Gaussian function.

\subsection{Boundary Motion}\label{sec:Boundary Motion}

The domain boundary is allowed to move either at constant speed $v_{\mathrm{s}}=\mathrm{const}$, or in oscillatory motion $v_{\mathrm{s}}=\Delta v_{\mathrm{s,a}}\cos \left(2\pi f_{\mathrm{s}} t\right)$, where $\Delta v_{\mathrm{s,a}}$ is the velocity amplitude of the spatial source motion, and $f_{\mathrm{s}}$ the corresponding frequency. With the initial boundary positions $X_0$ and $R_0$ in 1d and 3d spherical symmetry, respectively, the oscillatory motion is given by
\begin{equation}
X\left(t\right) = X_0 + \dfrac{\Delta v_{\mathrm{s,a}}}{2\pi f_{\mathrm{s}}}\sin\left(2\pi f_{\mathrm{s}}t\right)
\label{boundaryMotion}
\end{equation}
in 1d. In spherical coordinates, the same equation applies with $X$ and $X_0$ replaced by $R$ and $R_0$, respectively.

\subsection{Pressure Excitation}\label{sec:Pressure Excitation}

The sinusoidal pressure radiation at the wave emitting boundary is characterized by the frequency $f_0$ and the amplitude $\Delta p_{\mathrm{a}}$. The three following excitation functions are considered:
\begin{flalign}
\textnormal{Pulse excitation:} \hspace{0.3cm} &
p_{\mathrm{ex}}\left(t\right) = W\left(t\right) G\left(t\right)
\label{pulseExcitation} 
\\
\textnormal{Continuous excitation:} \hspace{0.3cm} &
 p_{\mathrm{ex}}\left(t\right) = Z\left(t\right) \mathcal{G}\left(t\right)
\label{contiExcitation2}
\\
 &
p_{\mathrm{ex}}\left(t\right) = W\left(t\right) \mathcal{G}\left(t\right)
\label{contiExcitation}
\end{flalign}
Pulse excitation (Eq. \eqref{pulseExcitation}) means that the pressure excitation involves only one wave period, whereas the continuous excitation (Eqs. \eqref{contiExcitation2} and \eqref{contiExcitation}) involves multiple wave periods, in this context also referred to as a wave train \citep{Blackstock_1966}. With $\Delta p_a$ being the maximum pressure of the periodic excitation and $f_0$ the corresponding frequency, Eqs. \eqref{pulseExcitation}, \eqref{contiExcitation2}, and \eqref{contiExcitation} are constructed from two different sinusoidal functions, given by
\begin{equation}
Z\left(t\right) = \Delta p_{\mathrm{a}}\sin\left(2\pi f_0 t\right),
\label{planeWave2}
\end{equation}
\begin{equation}
W\left(t\right) = \dfrac{\Delta p_{\mathrm{a}}}{2}\left[1-\cos\left(2\pi f_0 t\right)\right],
\label{planeWave}
\end{equation}
and two different envelopes $G\left(t\right)$ and $\mathcal{G}\left(t\right)$. The Gaussian envelope $G\left(t\right)$ is given by
\begin{equation}
G\left(t\right) = G_{\mathrm{ref}}^{4\left(f_0 t + \frac{1}{2} - N_{\mathrm{p}} \right)^2},
\hspace{0.3cm} \textnormal{where} \hspace{0.3cm}
N_{\mathrm{p}} \in \mathbb{N}^+.
\label{GaussEnv3}
\end{equation}
In Eq. \eqref{GaussEnv3}, $N_{\mathrm{p}}$ is the number of past wave periods of the functions $W\left(t\right)$ and $Z\left(t\right)$ at the peak incident of the envelope, and $G_{\mathrm{ref}}$ a reference value to adjust the steepness of the envelope. For $N_{\mathrm{p}}=1$, the peak of the envelope ($G=1$) coincides with the peak of the first excitation period at $t = 1/\left(2f_0\right)$ such that $\Delta p_{\mathrm{ex}} = \Delta p_{\mathrm{a}}$, while the discontinuity of the Laplacian at $t=0$ and $t=1/f_0$ is mitigated by the factor $G_{\mathrm{ref}}$. The start-up discontinuity can be further reduced by increasing $N_p$, which comes at the cost of additional computation time as $t = \left(N_p -1/2\right)/f_0$ marks the start of the excitation wave period of interest. In the present study, we choose $N_{\mathrm{p}}=10$. The Gaussian envelope $\mathcal{G}\left(t\right)$ is given by
\begin{equation}
\mathcal{G}\left(t\right) =
\left\{
\begin{array}{lll}
G\left(t\right) & \textnormal{for} & f_0 t \le N_{\mathrm{p}} - \frac{1}{2}
\\[10pt]
1 & \textnormal{for} & f_0 t > N_{\mathrm{p}} - \frac{1}{2}
\end{array}
\label{GaussConti}
\right.
\end{equation}
Hence, $\mathcal{G}\left(t\right)$ coincides with $G\left(t\right)$ up to the peak of the Gaussian profile. At the peak, $\mathcal{G}\left(t\right)$ is deactivated in order to continue with a continuous exciation. The evolution of the above excitation functions given by Eqs. \eqref{pulseExcitation}, \eqref{contiExcitation2}, and \eqref{contiExcitation} are depicted in Fig. \ref{fig:envFunSlope_contiPulse}. Eq. \eqref{pulseExcitation} represents an isolated pulse (black solid line in Fig. \ref{fig:envFunSlope_contiPulse}). Eqs. \eqref{contiExcitation2} and \eqref{contiExcitation} represent a continuous, periodic excitation. In Eq. \eqref{contiExcitation2}, the pressure oscillates around $p=0$ (blue solid line in Fig. \ref{fig:envFunSlope_contiPulse}), and in Eq. \eqref{contiExcitation}, the pressure oscillates around $\Delta p_{\mathrm{a}}/2$ (orange solid line in Fig. \ref{fig:envFunSlope_contiPulse}).

\subsection{Shape of the Gaussian Envelope}\label{sec:Shape of the Gaussian Envelope}

\begin{figure}[htb]
\begin{center}
{\includegraphics[width=0.7\linewidth]{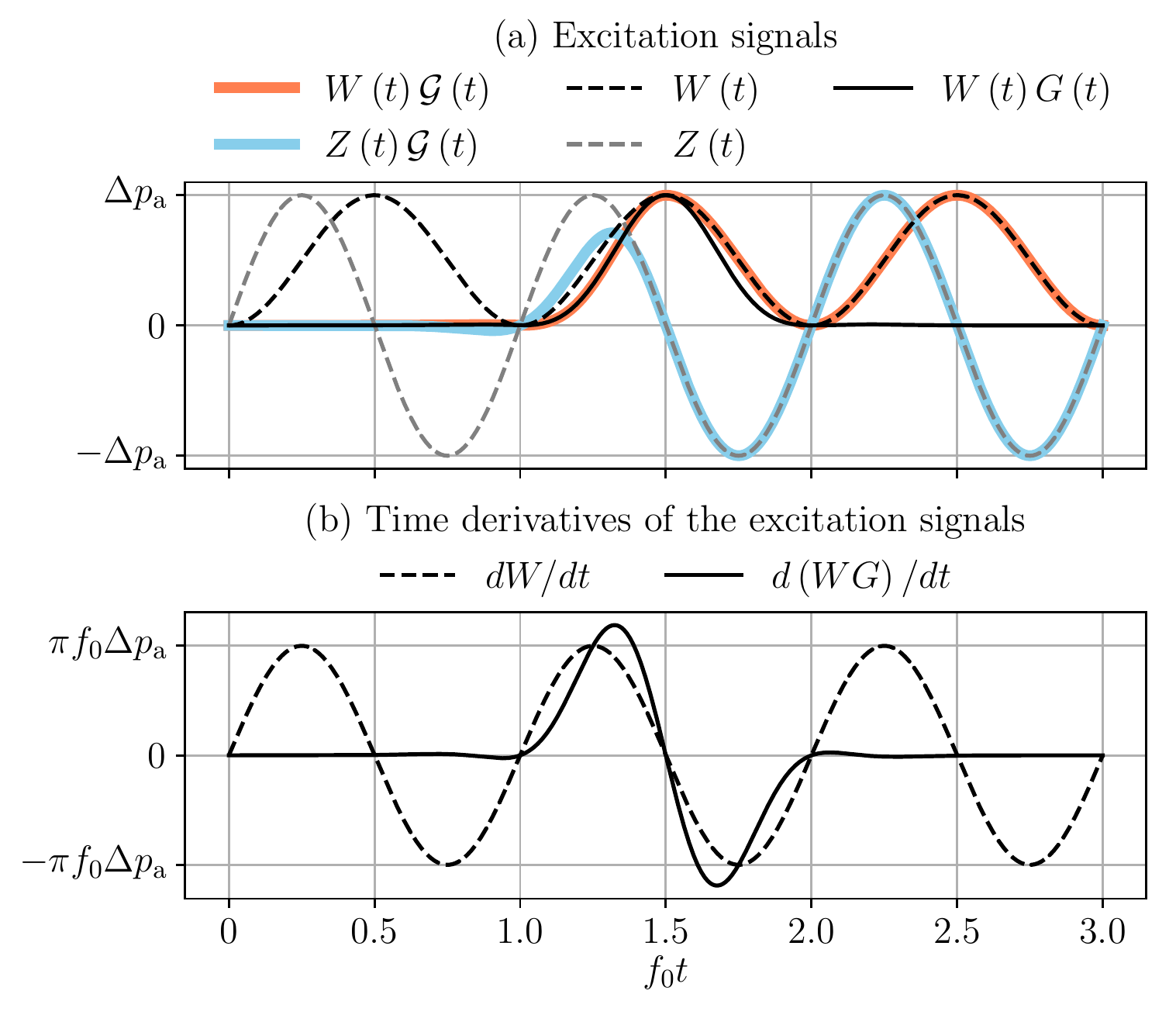}}
\caption{Gauss enveloped sinusoidal functions for pulse excitation (Eq. \eqref{pulseExcitation}) and continuous excitation (Eqs. \eqref{contiExcitation2} and \eqref{contiExcitation}). The maximum value of the derivative of the enveloped pulse function is larger than the one of the purely sinusoidal wave, which is accounted for by the factor $\alpha$ in the shock formation distance $x_{\mathrm{sh}}$ given by Eq. \eqref{shockFormationDistance2}. The top figure (a) depicts the excitation signals and the bottom figure (b) the corresponding time derivatives.}
\label{fig:envFunSlope_contiPulse}
\end{center}
\end{figure}

The choice of the reference value $G_{\mathrm{ref}}$ in Eq. \eqref{GaussEnv3} to adjust the shape of the Gaussian envelope affects the shock formation distance of the enveloped wave period. The shock formation distance in a nonlinear medium is given by \citep{Blackstock_et_al_1998}
\begin{equation}
x_{\mathrm{sh}} = \dfrac{\rho_0 c^3}{\beta \left(d p_{\mathrm{ex}}/d t\right)_{\mathrm{max}}},
\label{shockFormationDistance}
\end{equation}
where $\left(d p_{\mathrm{ex}}/d t\right)_{\mathrm{max}}$ is the maximum value of the time derivative of the excitation signal. For Eq. \eqref{planeWave}, we have $\left(d p_{\mathrm{ex}}/d t\right)_{\mathrm{max}} = \left(dW/dt\right)_{\mathrm{max}} = \pi \Delta p_{\mathrm{a}}f_0$ at $1/4$ of the wave period. This value can be recovered for the enveloped function at this particular time instant by imposing the condition
\begin{equation}
\left.\dfrac{\mathrm{d} p_{\mathrm{ex}}}{\mathrm{d} t}\right|_{t=\frac{1}{4f_0}} = \pi \Delta p_{\mathrm{a}}f_0
\hspace{0.3cm} \Rightarrow \hspace{0.3cm}
G_{\mathrm{ref}} = \dfrac{1}{\left(1-\dfrac{\ln\left(G_{\mathrm{ref}}\right)}{\pi}\right)^4},
\label{Gref}
\end{equation}
which holds for $G_{\mathrm{ref}}=1.5559$. However, with this value of $G_{\mathrm{ref}}$, $\left(d p_{\mathrm{ex}}/d t\right)_{\mathrm{max}}$ of Eq. \eqref{pulseExcitation} exceeds the one of the purely sinusoidal wave by a factor of $\alpha=1.1894$, which must be accounted for in the shock formation distance. Further taking the Doppler shift in the wavelength according to the linear dispersion relation into account, the adjusted shock formation distance becomes
\begin{equation}
x_{\mathrm{sh}} = \dfrac{\rho_0 c_0^2\left(c_0 + v_s\right)}{\pi \alpha \beta \Delta p_{\mathrm{a}}f_0}.
\label{shockFormationDistance2}
\end{equation}
We emphasize that due to the choice of the Gaussian envelope for the pulse excitation, some small remainders of the two neighboring wave periods are visible in the corresponding wave profile (see the black solid lines in Fig. \ref{fig:envFunSlope_contiPulse}), which are also present in the numerical solution.

\section{Verification}\label{sec:Verification}

In this section, the accuracy and convergence properties of the time-explicit finite difference method for solving Eq. \eqref{LinEqn} are investigated by comparing the numerical solutions with analytical reference solutions for a hierarchy of test cases. For the reasons given in Sec. \ref{sec:Introduction}, the sound diffusivity is neglected, meaning that $\delta = 0$ in Eq. \eqref{Westervelt_eqn}. First, the fluid parameters and the numerical base settings are specified. Then, the results of the verification study are presented. An in-depth discussion of the results and observations is given in the following section Sec. \ref{sec:Discussion}.

\subsection{Fluid Properties and Wave Parameters}\label{sec:Fluid Properties}

The verification study involves several exercises, where the reference wavelength, sound speed and frequency of the linear wave are $\lambda_0=15 \: \mathrm{mm}$, $c_0=1500 \: \mathrm{m/s}$, and $f_0 = c_0/\lambda_0 = 100 \: \mathrm{kHz}$. The reference density $\rho_0 = 1000 \: \mathrm{kg/m^3}$ is assumed to be constant. The wave amplitude in the excitation functions given by Eqs. \eqref{planeWave2} and \eqref{planeWave} is chosen to be $\Delta p_{\mathrm{a}}=10 \: \mathrm{MPa}$, where the effective amplitude of Eq. \eqref{planeWave} is half of the amplitude of Eq. \eqref{planeWave2}. The results will be presented in terms of dimensionless quantities. Nevertheless, it is important to mention that the choice of the wave geometry and fluid parameters affects the distortion of the nonlinear wave and in particular the shock formation distance given by Eq. \eqref{shockFormationDistance2}. This is illustrated by assuming a purely sinusoidal wave ($\alpha=1$ in Eq. \eqref{shockFormationDistance2}) and a fixed wave emitting boundary ($v_{\mathrm{s}}=0$). Further applying the dispersion relation $c_0=\lambda f$, Eq. \eqref{shockFormationDistance2} can be rewritten as
\begin{equation}
x_{\mathrm{sh}} =\dfrac{1}{\pi} \dfrac{\lambda}{\Delta p_{\mathrm{a}}} \dfrac{\rho_0 c_0^2}{\beta}.
\label{shockFormationDistance3}
\end{equation}
The factor $\lambda/\Delta p_{\mathrm{a}}$ in Eq. \eqref{shockFormationDistance3} represents the influence of the wave geometry on the shock formation distance, whereas the factor $\rho_0 c_0^2/\beta$ represents the effect of the fluid properties on the rate of nonlinear wave distortion. The inverse of $\rho_0 c_0^2/\beta$ is the coefficient of the nonlinear term in the Westervelt equation given by Eq. \eqref{Westervelt_eqn}. Hence, geometrically similar waves are obtained for any parameter configuration for which $\lambda/\Delta p_{\mathrm{a}}=\mathrm{const}$ and $\rho_0 c_0^2/\beta=\mathrm{const}$.

\subsection{Numerical Settings}\label{sec:Numerical Settings}

The $\mathrm{CFL}$ number is based on the small signal sound speed $c_0$ in Eq. \eqref{Westervelt_eqn}. When the computational domain is compressed due to the motion of the wave emitting boundary, it is based on the initial step size $\Delta x_0$ and given by
\begin{equation}
\mathrm{CFL}_0 = \dfrac{c_0 \Delta t}{\Delta x_0}.
\label{CFL}
\end{equation}
For $\Delta x = \mathrm{const}$, we have $\mathrm{CFL} = \mathrm{const} = \mathrm{CFL}_0$. The base settings for the spatial resolution and the temporal resolution are $N_{\mathrm{ppw}}=320$ points per initial wavelength $\lambda_0$ and $\mathrm{CFL}_0=0.1$, respectively. The value of the relaxation factor $\gamma$ in Eq. \eqref{LinEqn_discr_corr}, representing the weight of the corrected solution relative to the predicted solution in the predictor-corrector method, is $\gamma=0.5$. These settings are used if not mentioned otherwise, and they are only changed for the purpose of accuracy and convergence studies.

\subsection{Linear Doppler Modulation of the Linear Wave}\label{sec:Linear Doppler Modulation of the Linear Wave}

\begin{figure}[htb]
\begin{center}
{\includegraphics[width=0.8\linewidth]{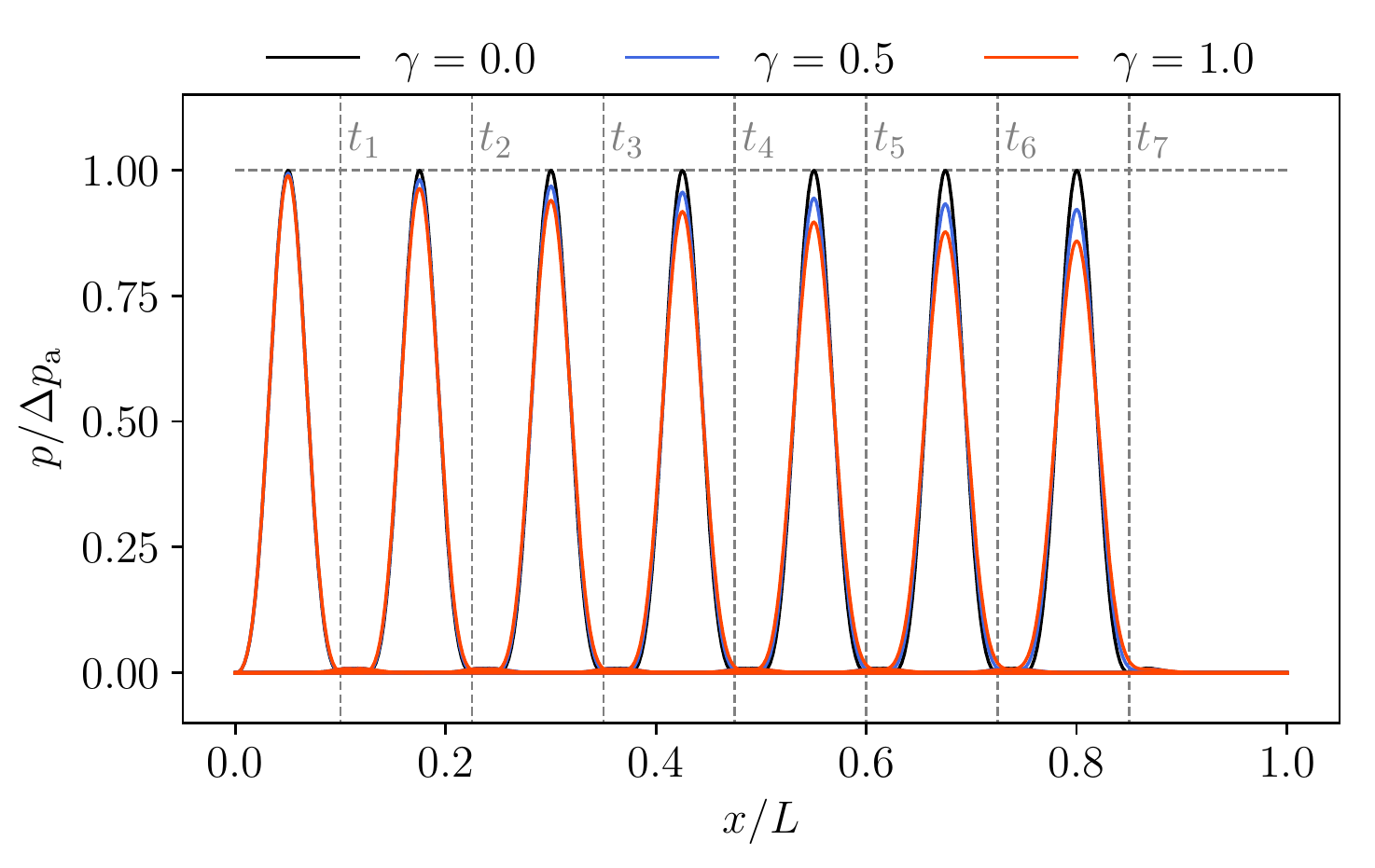}}
\caption{Pressure wave profiles obtained from the pulse excitation (Eq. \eqref{pulseExcitation}) for different values of the relaxation parameter $\gamma$, where $\gamma=0$ corresponds to the standard FTCS scheme. The spatial resolution is $N_{\mathrm{ppw}}=80$ grid points per wavelength $\lambda_0$. The wave emitting boundary is at rest ($v_{\mathrm{s}}=0$) and the medium is linear ($\beta=0$). The dashed vertical lines indicate the wave front positions based on $c_0$ at the time instants $t_1$ to $t_7$ at a spacing of $1.25/f_0$.}
\label{fig:correctorWeight}
\end{center}
\end{figure}

First, the effect of the anti-dispersive predictor-corrector method is demonstrated in Fig. \ref{fig:correctorWeight} for a linear wave ($\beta=0$ in Eq. \eqref{Westervelt_eqn}) on a static grid ($v_{\mathrm{s}}=0$) and for different values of the relaxation factor $\gamma$ in Eq. \eqref{LinEqn_discr_corr}. The solution is depicted for a spatial resolution of $N_{\mathrm{ppw}}=80$ points per wavelength $\lambda_0$. Hence, a lower resolution as compared to the base case is chosen to demonstrate the effect of the relaxation factor $\gamma$. The propagation of a single wave period is investigated, obtained from the pulse excitation function given by Eq. \eqref{pulseExcitation}. The dashed vertical lines indicate the wave front positions at the time instants $t_1$ to $t_7$ (spacing of $1.25/f_0$) predicted from the constant wave propagation speed $c_0$. For $\gamma=0$, the method corresponds to the standard FTCS scheme. For increasing $\gamma$, the corrector step in Eq. \eqref{LinEqn_discr_corr} is given increasing weight. It can be seen that the wave experiences an increasing amount of diffusion as $\gamma$ increases, where the amplitude decreases while the wavelength slightly increases. With the standard FTCS scheme, both the amplitude and the wavelength are preserved perfectly for the given resolution. However, the FTCS solution is also very sensitive to numerical errors, which will unavoidably occur if the wave is nonlinear and/or if grid deformation is involved. A weight of $\gamma=0.5$ is chosen for the following cases, not only to compromise between numerical stability and numerical diffusion, but also because the predictor method by \citet{Dey_and_Dey_1983} is second order accurate in time for $\gamma=0.5$ when applied to an ordinary differential equation of the form $\dot y = f\left(y\right)$ \citep{Dey_1999}, even though in the present work, second order accuracy is not expected due the presence of the mixed derivative in Eq. \eqref{LinEqn} and/or the formation of shocks.

\begin{figure}[htb]
\begin{center}
{\includegraphics[width=0.85\linewidth]{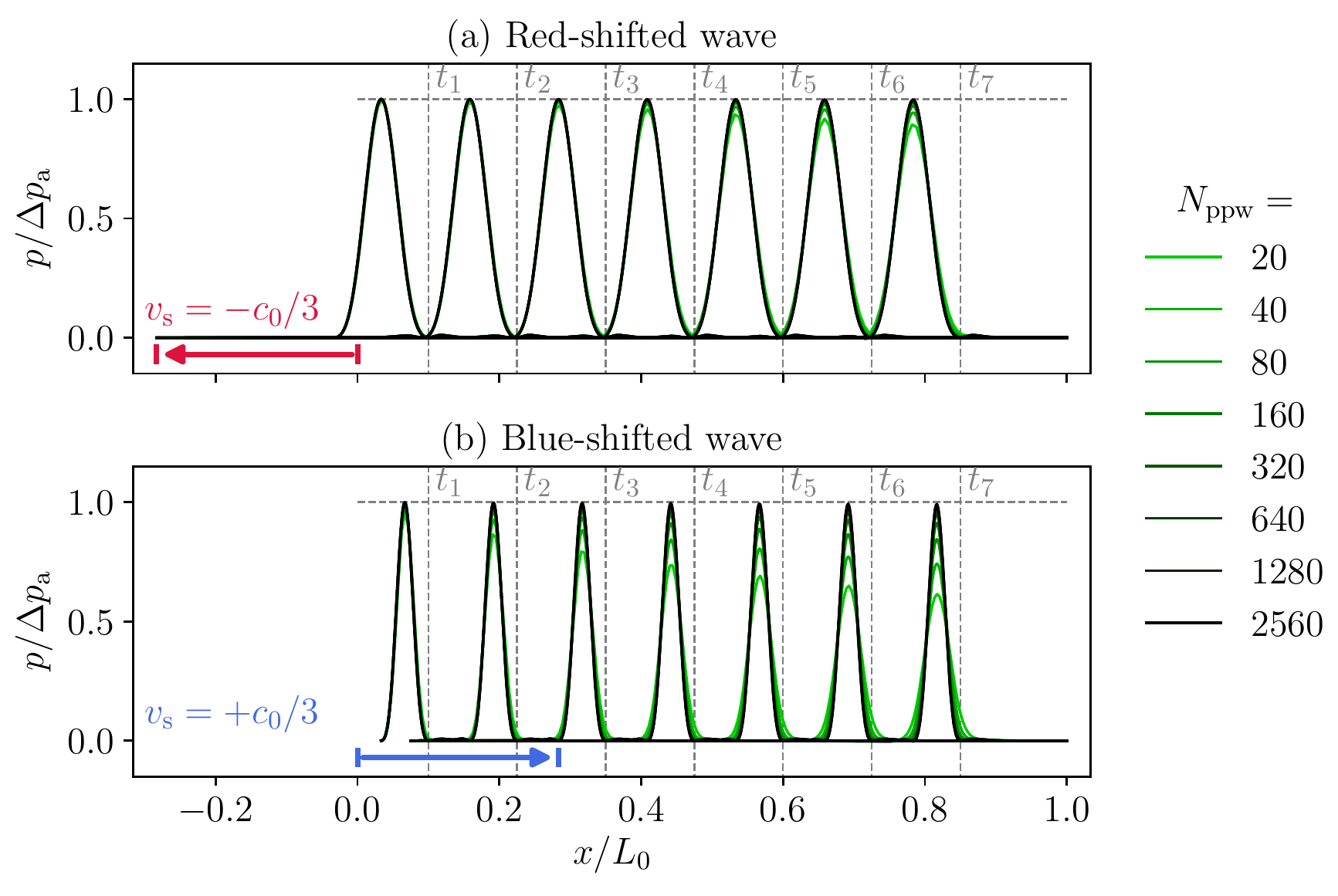}}
\caption{Pressure wave profiles obtained from the pulse excitation (Eq. \eqref{pulseExcitation}). $N_{\mathrm{ppw}}$ is the number of discretization points per wavelength $\lambda_0$. The wave is excited at the left boundary, which is moving in the negative $x$-direction in the top figure (a) (red-shift), and in the positive $x$-direction in the bottom figure (b) (blue-shift). The medium is linear ($\beta=0$). The dashed vertical lines indicate the wave front positions based on $c_0$ at the time instants $t_1$ to $t_7$ at a spacing of $1.25/f_0$.}
\label{fig:dxStudy_redblueShift}
\end{center}
\end{figure}

Fig. \ref{fig:dxStudy_redblueShift} shows the spatial wave profiles at different time instants $t_1$ to $t_7$ (spacing of $1.25/f_0$) for the boundary moving to the left and to the right, respectively. For $v_{\mathrm{s}}<0$, the excitation source moves opposite to the wave propagation direction, which causes a stretching (red shift) of the wavelength $\lambda_0$, whereas for $v_{\mathrm{s}}>0$, the source moves along the direction of wave propagation, such that the wave profile is compressed accordingly (blue shift). The wave profiles are shown for different grid densities. The other numerical settings correspond to the base settings specified in Sec. \ref{sec:Numerical Settings}. Due to the stretching/compression of the physical domain, the $\mathrm{CFL}$-number, initially being equal to 0.1, changes over time. The corresponding final $\mathrm{CFL}$ numbers at $t_7$ are $0.072$ for the red-shifted wave and $0.140$ for the blue-shifted wave. Due to the linearity of the wave, the source motion relative to the resting medium is assumed to have no effect on the wave amplitude and, since $v_{\mathrm{s}} < c_0$, also the position of the wave front must remain unaffected. The predicted Doppler wavelengths are $\lambda_0 \pm v_{\mathrm{s}}/f_0 = \left\{20; \: 10\right\} \: \mathrm{mm}$. It can bee seen in Fig. \ref{fig:dxStudy_redblueShift} that the wavelengths are in good agreement with the prediction, while the position of the wave front, as indicated by the vertical dashed lines at $t_1$ to $t_7$ (spacing of $1.25/f_0$), remains unaffected by the boundary motion. Also, the amplitude loss and smearing of the wave profile decreases with increasing spatial resolution.

\begin{figure}[htb]
\begin{center}
{\includegraphics[width=0.7\linewidth]{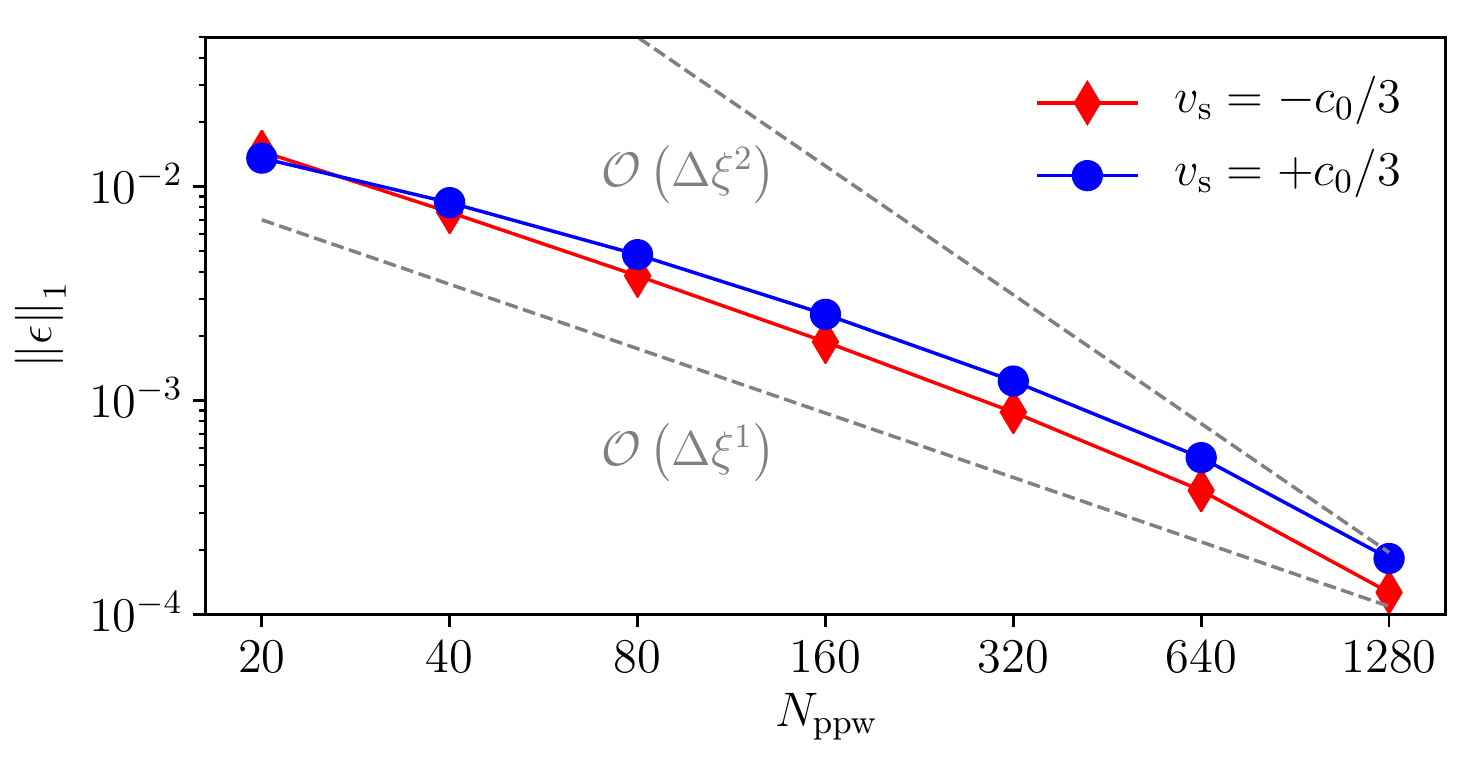}}
\caption{Evolution of the error $\left\|\epsilon\right\|_1$ defined by Eq. \eqref{L1eps} over $N_{\mathrm{ppw}}$ for the red-shifted and the blue-shifted waves in Fig. \ref{fig:dxStudy_redblueShift}. $N_{\mathrm{ppw}}$ is the number of grid points per wavelength $\lambda_0$ of the unmodulated wave ($v_{\mathrm{s}}=0$). The gray dashed lines indicate the slopes corresponding to a first-order and a second-order convergence, respectively.}
\label{fig:convDiagDoppler}
\end{center}
\end{figure}

Fig. \ref{fig:dxStudy_redblueShift} suggests that for the same initial spatial resolution, the profile of the blue-shifted wave undergoes more numerical diffusion than the red-shifted wave. In order to investigate this behavior in more detail, the relative error 
\begin{equation}
\epsilon\left(x,t\right) = \left\|p\left(x,t\right) - p_{2560}\left(x,t\right)\right\|
\label{epsilom}
\end{equation}
is introduced, which measures the deviation of the pressure $p\left(x\right)$ for a given spatial resolution relative to the pressure $p_{2560}$ obtained for the finest grid, for which the initial number of grid points per wavelength $\lambda_0$ ($v_{\mathrm{s}}=0$) is $N_{\mathrm{ppw}}=2560$. Hence, $\epsilon = 0$ for the finest grid. Given the instantaneous domain boundaries $X\left(t\right)$ and $L$, the $L_1$ norm of $\epsilon$ at time $t$ is given by
\begin{equation}
\left\|\epsilon\left(t\right)\right\|_1
=
\dfrac{ 1}{\lambda_0 \Delta p_{\mathrm{a}} N_{\mathrm{p}}} \left(\dfrac{f_0}{f}\right)^2  \int_{X\left(t\right)}^{L}\epsilon\left(x,t\right)dx.
\label{L1eps}
\end{equation}
The error is normalized by the excitation pressure amplitude $\Delta p_{\mathrm{a}}$. The integral in Eq. \eqref{L1eps} is carried out over the entire domain length, but normalized by the reference wavelength $\lambda_0$ because the domain only contains one single pulse due to the pulse excitation given by Eq. \eqref{pulseExcitation}. $N_{\mathrm{p}}=7.5$ is the number of wavelengths $\lambda_0$ that an unmodulated wave ($v_{\mathrm{s}}=0$) would have traveled until time $t$, and accounts for the fact that the error of the wave profile increases over time. The factor $f_0/f$ is the ratio of frequencies between the unmodulated and the Doppler-shifted waves. The square of this factor accounts for the fact that due to the linear Doppler shift in the wavelength, both the effective spatial resolution of the wave profile and the number of traveled wavelengths at a given time instant change accordingly. Fig. \ref{fig:convDiagDoppler} shows the resulting evolution of $\left\|\epsilon\right\|_1$ over $N_{\mathrm{ppw}}$ for the red-shifted and the blue-shifted waves in Fig. \ref{fig:dxStudy_redblueShift}. It can be seen that both profiles converge at very similar rates when the Doppler related changes of the effective spatial resolution and the number of traveled wavelengths are taken into account. For reference, the gray dashed slopes represent $\mathcal{O}\left(\Delta \xi^1\right)$ and $\mathcal{O}\left(\Delta \xi^2\right)$ convergence, respectively. Both wave profiles show first order convergence for the medium grids. The order of convergence is somewhat smaller than the one for the coarse grids, and it is close to $\mathcal{O}\left(\Delta \xi^2\right)$ for the finest grid.

\subsection{Nonlinear Doppler Modulation of the Linear Wave}\label{sec:Nonlinear Doppler Modulation of the Linear Wave}

\begin{figure}[htb]
\begin{center}
{\includegraphics[width=0.75\linewidth]{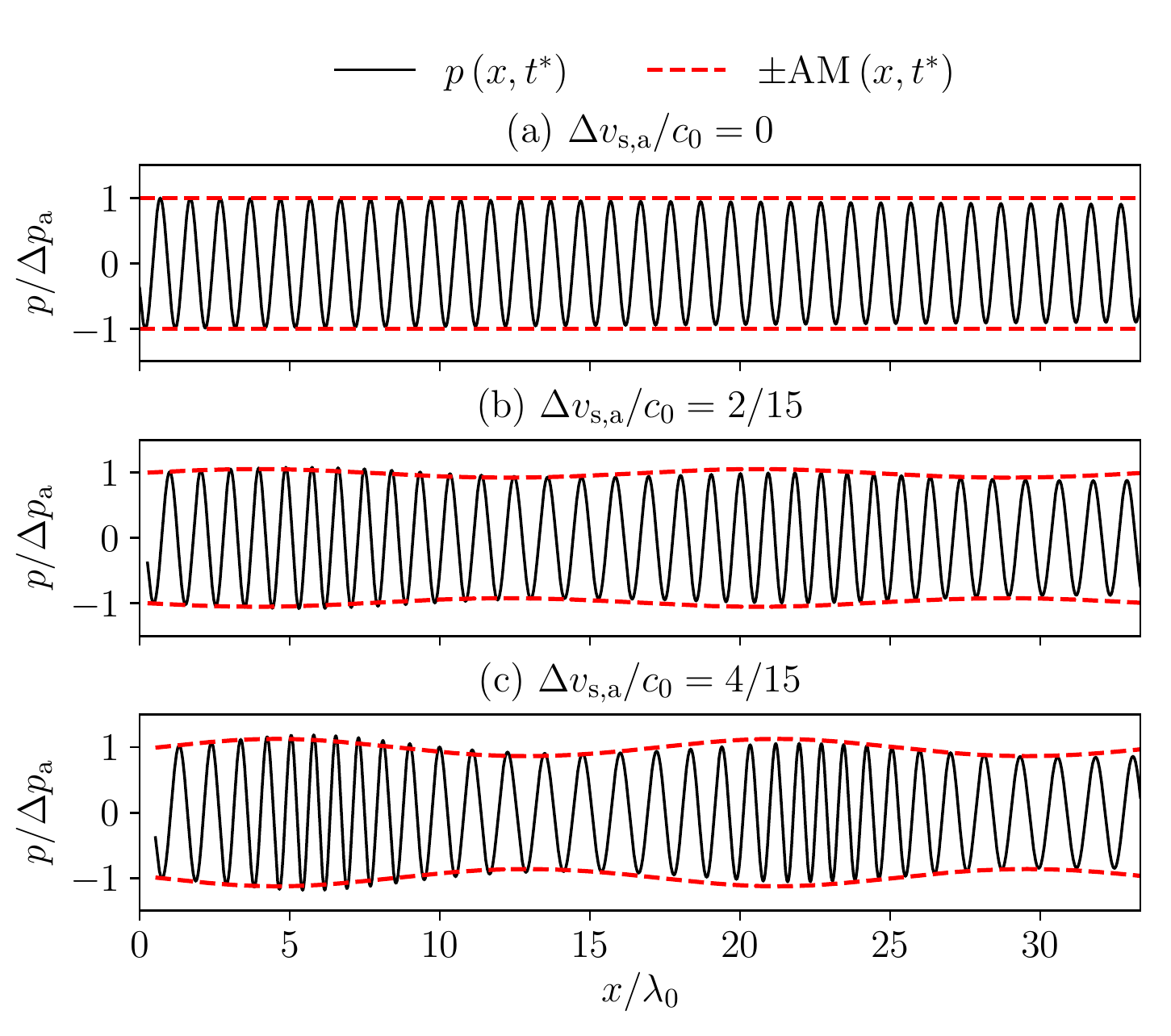}}
\caption{Instantaneous wave profile obtained from continuous excitation (Eq. \eqref{contiExcitation2}) at the left boundary, which is in an oscillatory motion around $X_0 = 0$ according to Eq. \eqref{boundaryMotion}, where $f_{\mathrm{s}}/f_0=3/50$. The velocity amplitude $\Delta v_{\mathrm{s,a}}$ is varied systematically in the sub-figures (a), (b), and (c). The pressure wave profiles and the amplitude modulation factors $\mathrm{AM}$ given by Eq. \eqref{AM} \citep{Christov_2017} are evaluated at $f_0 t = 38.44$ after the start of the acoustic excitation. The medium is linear ($\beta=0$).}
\label{fig:tremolo}
\end{center}
\end{figure}

\begin{figure}[htb]
\begin{center}
{\includegraphics[width=0.75\linewidth]{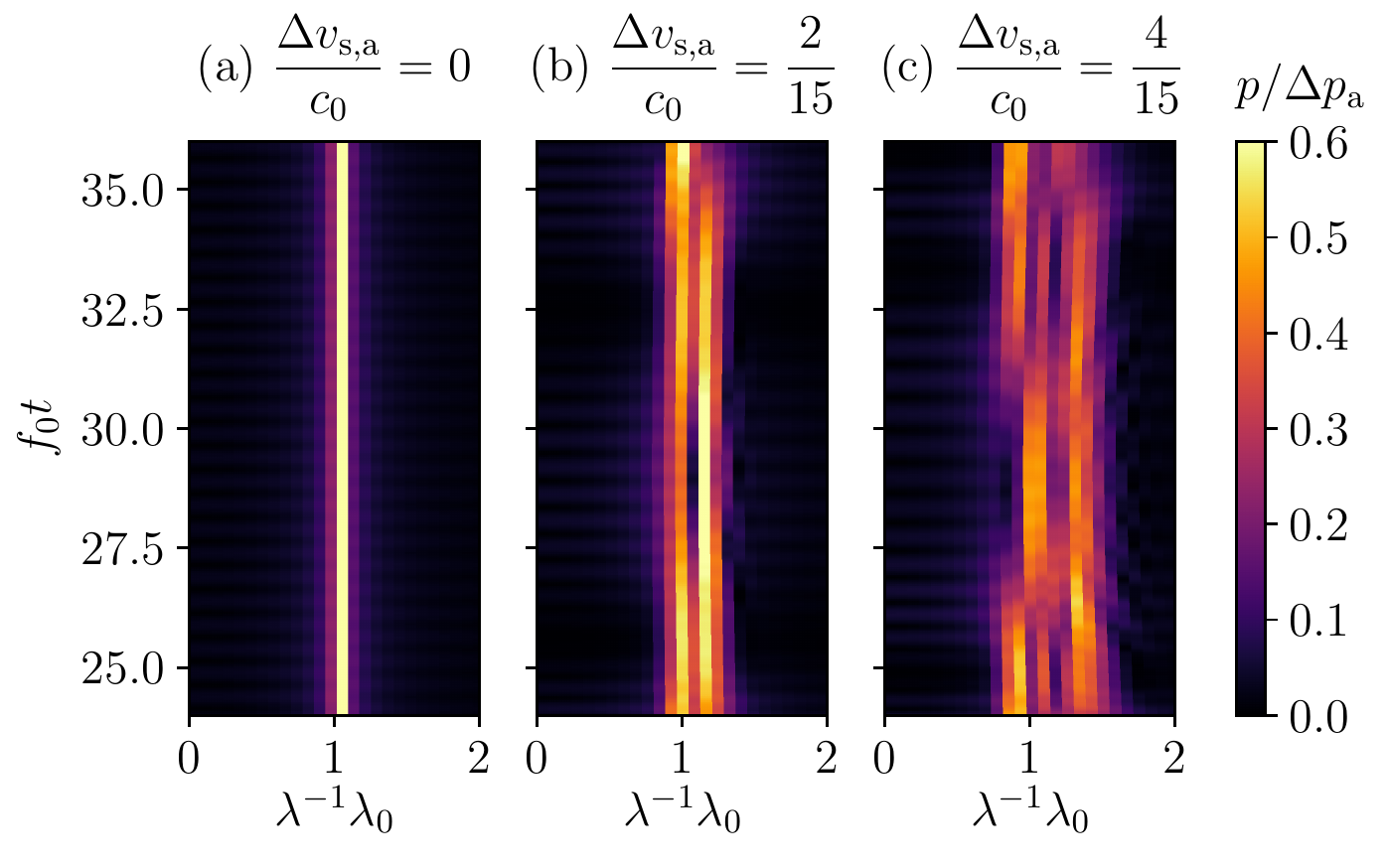}}
\caption{Spatial pressure amplitude spectra for the waves in Fig. \ref{fig:tremolo}, where the velocity amplitude $\Delta v_{\mathrm{s,a}}$ is varied systematically in the sub-figures (a), (b), and (c), and where $\lambda_0$ is the wavelength associated with a fixed wave emitting boundary. The spectra are evaluated in the left half of the computational domain, which has been passed by the front of the wave train at the first time instant of the depicted time range.}
\label{fig:frequ_v0_v200_v400}
\end{center}
\end{figure}

\begin{figure}[htb]
\begin{center}
{\includegraphics[width=0.75\linewidth]{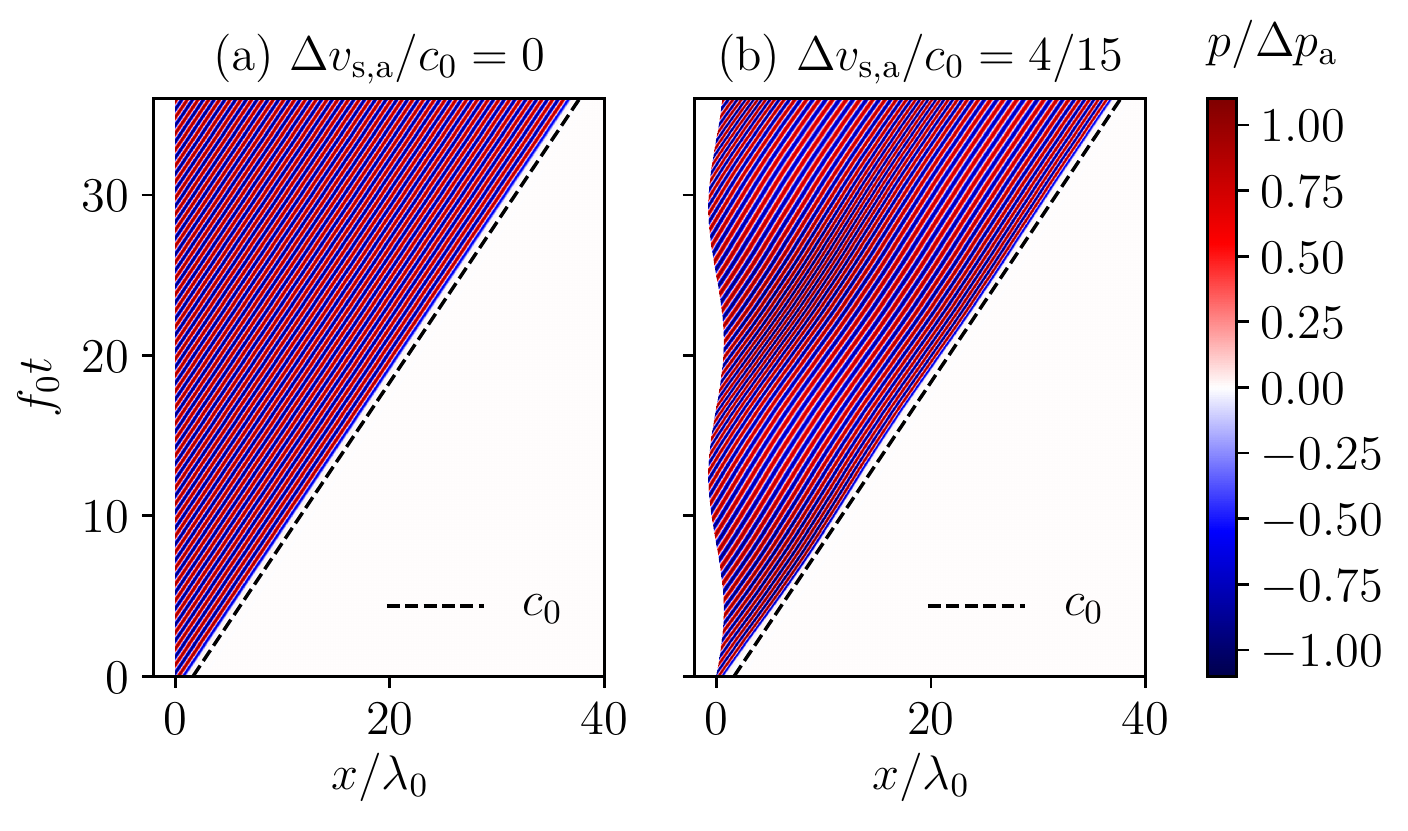}}
\caption{Pressure distributions in the $x$-$t$ plane obtained for the cases $\Delta v_{\mathrm{s,a}}/c_0=0$ (a) and $\Delta v_{\mathrm{s,a}}/c_0=4/15$ (b) in Fig. \ref{fig:tremolo}, where $f_{\mathrm{s}}/f_0=3/50$. The dashed black line indicates the direction of propagation in the $x$-$t$ plane associated with a constant propagation speed of $c_0$.}
\label{fig:xtPlot_tremolo_v0_v400}
\end{center}
\end{figure}

In the following, the continuous excitation function given by Eq. \eqref{contiExcitation2} is employed. Hence, the pressure oscillates symmetrically around $p=0$, which is also the case in the studies by \citet{Christov_2017} and by \citet{Gasperini_et_al_2021}. Fig. \ref{fig:tremolo} shows the wave profile for an oscillating wave emitting boundary. The frequency of the sinusoidal boundary motion is $f_{\mathrm{s}}=\left(3/50\right)f_0$, hence significantly smaller than the frequency of the propagating acoustic wave radiation. The resulting wave profiles are shown for three different velocity amplitudes of the moving source. It can be seen that different from the linear Doppler effect, the wave amplitude undergoes an amplitude modulation due to the acceleration of the wave emitting boundary. Making use of multiple-scales expansion, \citet{Christov_2017} presented an analytical solution of the linear wave equation subject to the periodic excitation at a moving boundary under the prerequisite of slowly varying coefficients. In the limit $f_{\mathrm{s}}\ll f_0$, \citet{Christov_2017} derived the wave amplitude modulation factor as
\begin{equation}
\begin{array}{lll}
\mathrm{AM}\left(x,t\right)
&=&
\exp\left\{-\dfrac{1}{2}\dfrac{\Delta v_{\mathrm{s,a}}}{c_0}\left[\cos\left(\omega_{\mathrm{s}} \left(t + \dfrac{x}{c_0}\right)-2\dfrac{\Delta v_{\mathrm{s,a}}}{c_0}\sin\left(\omega_{\mathrm{s}}t\right)\right)\right.\right. \\
&& - 
\left. \left. \cos\left(\omega_{\mathrm{s}}t-\dfrac{\Delta v_{\mathrm{s,a}}}{c_0}\sin\left(\omega_{\mathrm{s}}t\right)\right)\right]\right\}
\end{array}
\label{AM}
\end{equation}
for a sinusoidal boundary motion at the angular frequency $\omega_{\mathrm{s}}=2\pi f_{\mathrm{s}}$. Applying the amplitude modulation factor at $f_0 t=38.44$ after the start of the acoustic excitation gives the envelopes indicated by the red dashed lines in Fig. \ref{fig:tremolo}. The pressure wave profiles are evaluated at the same time instant. It is observed that the pressure amplitudes increasingly deviate from the waveform predicted by the asymptotic amplitude modulation factor as the velocity amplitude $\Delta v_{\mathrm{s,a}}/c_0$ increases. Fig. \ref{fig:frequ_v0_v200_v400} shows the evolution of the pressure amplitude over time and over the spatial frequency $\lambda^{-1}$, where $\lambda^{-1}$ is multiplied by the fundamental frequency $\lambda_0$. The spectra are computed for the left half of the domain, which has been passed over by the front of the wave train at the first time instant of the depicted time range. When $\Delta v_{\mathrm{s,a}}=0$, only the inverse of the fundamental wave frequency is visible, so that $\lambda^{-1}\lambda_0=1$. With increasing velocity amplitude $\Delta v_{\mathrm{s,a}}$, a broadband spectrum develops, which is predicted by \citet{Gasperini_et_al_2021} as well. Fig. \ref{fig:xtPlot_tremolo_v0_v400} shows the pressure distribution in the $x$-$t$ plane, where the black dashed line indicates the direction of wave propagation at speed $c_0$. For $\Delta v_{\mathrm{s,a}}=0$, the wave front exactly follows the path given by $c_0$, whereas for $\Delta v_{\mathrm{s,a}}/c_0=4/15$, small oscillations of the wave propagation speed are observed. The latter effect is predicted by \citet{Christov_2017}.

\subsection{Linear Doppler Modulation of the Nonlinear Wave}\label{sec:Linear Doppler Modulation of the Nonlinear Wave}

\begin{figure}[htb]
\begin{center}
{\includegraphics[width=0.85\linewidth]{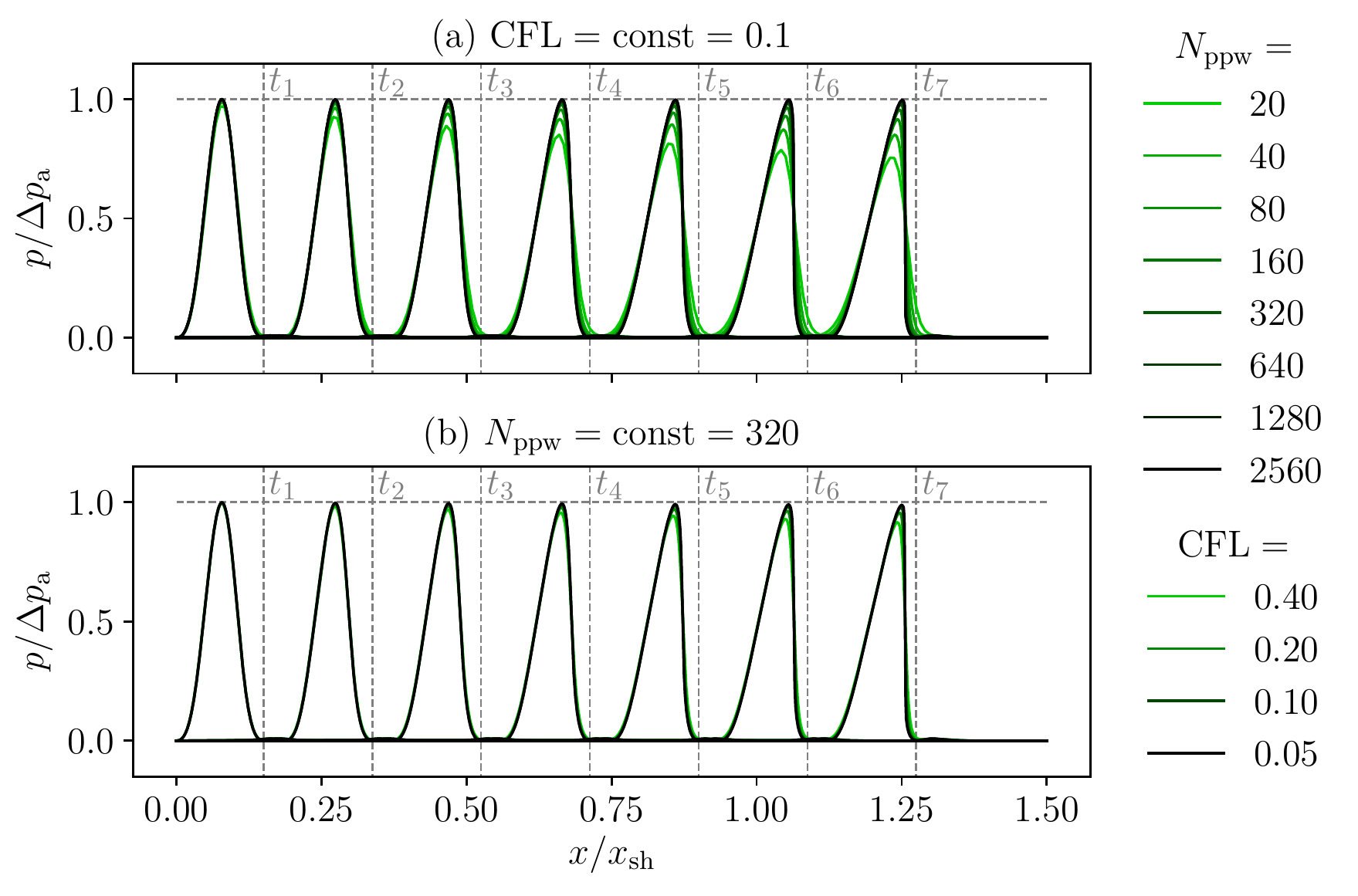}}
\caption{Pressure wave profiles obtained from the pulse excitation (Eq. \eqref{pulseExcitation}) for different spatial (a) and temporal (b) resolutions, where $\beta=9.0$ and $\gamma=0.5$. The wave emitting boundary is at rest ($v_{\mathrm{s}}=0$). $N_{\mathrm{ppw}}$ is the number of grid points per wavelength $\lambda_0$. The dashed vertical lines indicate the wave front positions based on $c_0$ at the time instants $t_1$ to $t_7$ at a spacing of $1.25/f_0$.}
\label{fig:dxdtStudy_shock}
\end{center}
\end{figure}

\begin{figure}[htb]
\begin{center}
{\includegraphics[width=0.8\linewidth]{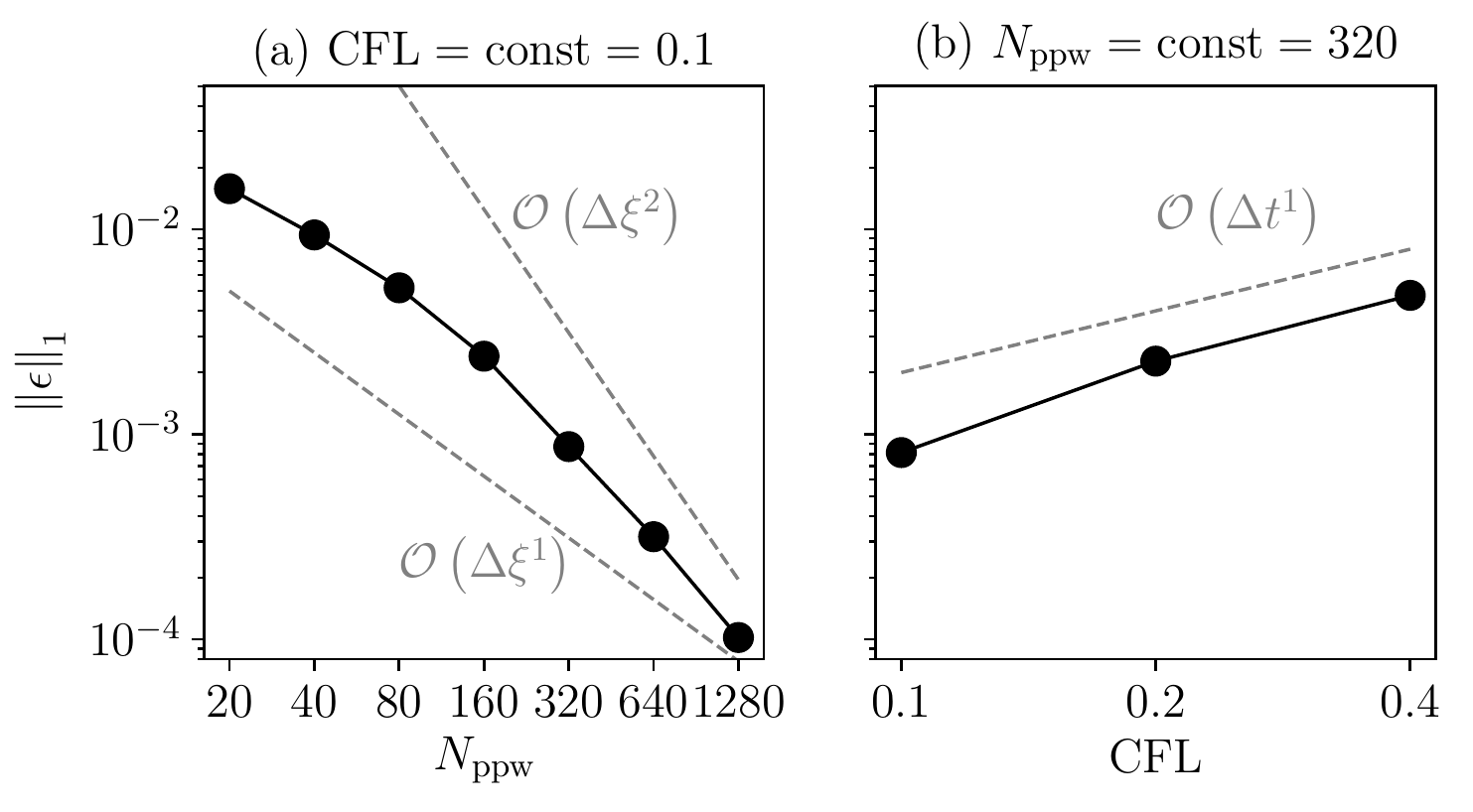}}
\caption{Evolution of the error $\left\|\epsilon\right\|_1$ defined by Eq. \eqref{L1eps} over $N_{\mathrm{ppw}}$ for a fixed $\mathrm{CFL}$ number (a) and over $\mathrm{CFL}$ for a fixed spatial resolution (b) for the nonlinear wave in Fig. \ref{fig:dxdtStudy_shock}. $N_{\mathrm{ppw}}$ is the number of grid points per wavelength $\lambda_0$. The gray dashed lines indicate the slopes corresponding to a first-order and a second-order convergence, respectively.}
\label{fig:convDiagShock}
\end{center}
\end{figure}

In the following, we assume a shock formation distance of $x_{\mathrm{sh},0}/\lambda_0 = 20/3$ for $v_{\mathrm{s}}=0$. For the given excitation pressure amplitude, this holds for $\beta=9.0$ in Eq. \eqref{shockFormationDistance2}. Fig. \ref{fig:dxdtStudy_shock} shows the wave profile at the time instants $t_1$ to $t_7$ (spacing of $1.25/f_0$) for different spatial resolutions at $\mathrm{CFL}=0.1$ (a), and for different $\mathrm{CFL}$ numbers at fixed spatial resolution (b). The wave emitting boundary is at rest ($v_{\mathrm{s}}=0$). Again, the wave amplitude exhibits convergent behavior with increasing spatial resolution, whereas the wave profile undergoes numerical diffusion on the coarser grids. As the wave approaches the shock formation distance $x_{\mathrm{sh}}$, the wave front steepens due to the nonlinearity of the medium. At $x_{\mathrm{sh}}$, a shock front starts to form. It is observed that the numerical diffusion on the coarser grids also mitigates the steepening of the wave profile, which is intended in order to provide numerical stability. A similar trend is observed for the variation of the $\mathrm{CFL}$ number, where the larger values of the $\mathrm{CFL}$ number result in a slight smearing of the wave profile and a slight loss of amplitude as well. Doubling the largest time step size once more ($\mathrm{CFL}=0.8$) results in an unstable solution. Analogously to Fig. \ref{fig:convDiagDoppler}, Fig. \ref{fig:convDiagShock} shows the evolution of the error $\left\|\epsilon\right\|_1$, given by Eq. \eqref{L1eps}, for the wave profiles depicted in Fig. \ref{fig:dxdtStudy_shock} over $N_{\mathrm{ppw}}$ (a) and over $\mathrm{CFL}$ (b). Regarding the spatial resolution, it is observed that the convergence rates in Figs. \ref{fig:convDiagDoppler} and \ref{fig:convDiagShock} are very similar. However, in Fig. \ref{fig:convDiagShock} the wave profile approaches second order convergence at a somewhat lower resolution than in Fig. \ref{fig:convDiagDoppler}. The slightly better convergence behavior in Fig. \ref{fig:convDiagShock} as compared to Fig. \ref{fig:convDiagDoppler} may be explained by the circumstance that the kinematic coupling term $\mathcal{K}$ given by Eq. \eqref{kinCoupling}, which involves the mixed derivative and which appears on the right-hand side of the discretized equation (see Eq. \eqref{LinEqn_discr}), is zero in Fig. \ref{fig:convDiagShock}. Regarding the temporal resolution in Fig. \ref{fig:convDiagShock}, the convergence rate is approximately of order one as well.
 
\begin{figure}[htb]
\begin{center}
{\includegraphics[width=\linewidth]{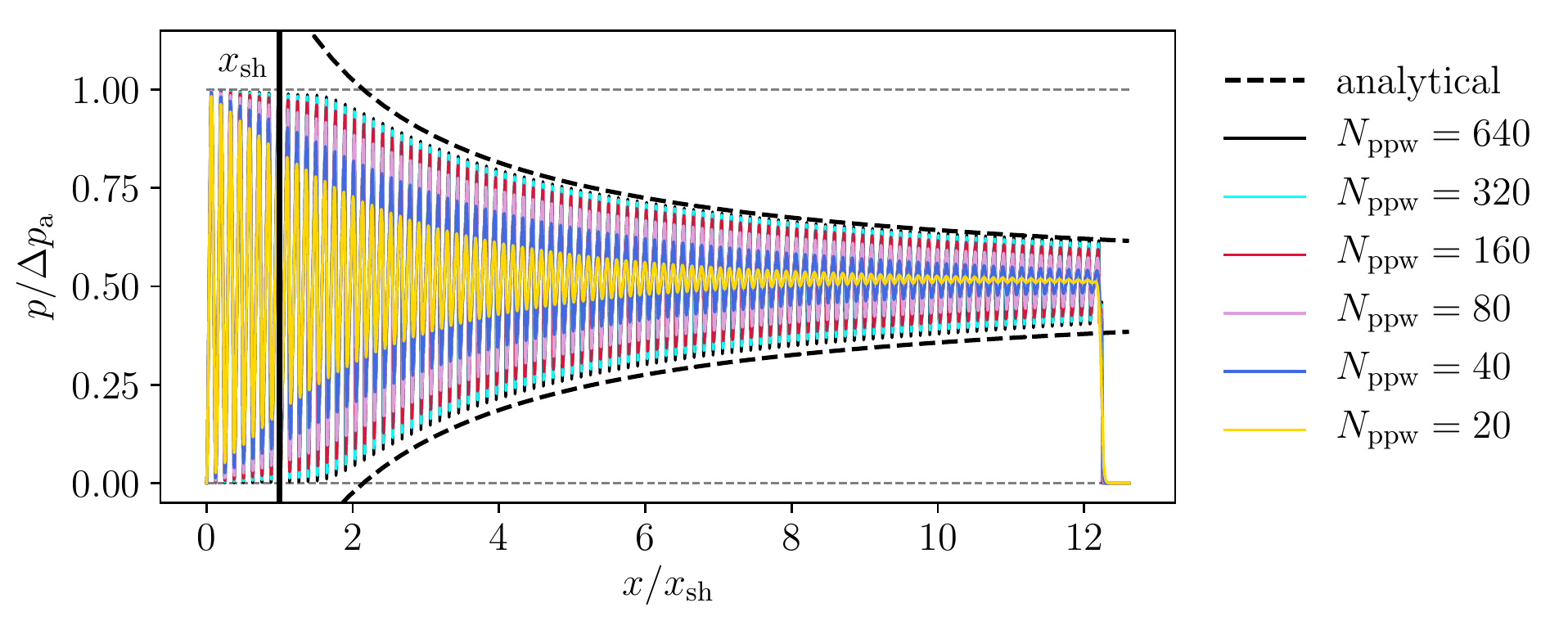}}
\caption{Pressure wave profile at $t=96/f_0$ for the continuous excitation function given by Eq. \eqref{contiExcitation} and $\beta=9.0$ for different spatial resolutions, where $N_{\mathrm{ppw}}$ is the number of grid points per wavelength $\lambda_0$. The analytical solution represents the asymptotic decay of the shock amplitude for large $x/x_{\mathrm{sh}}$ as given by Eq. \eqref{ampDecay} \citep{Blackstock_1966}, where $\Delta \phi_{\mathrm{a}}=\Delta p_{\mathrm{a}}/2$.}
\label{fig:longSimu_dxStudy}
\end{center}
\end{figure}

In order to investigate the propagation behavior of the shock in more detail, Fig. \ref{fig:longSimu_dxStudy} shows the instantaneous wave profile at $t=96/f_0$ for the continuous excitation function given by Eq. \eqref{contiExcitation}, where the wave travel distance involves a little more than 12 shock formation distances. The results are shown for different spatial resolutions. For low spatial resolution, a pronounced decay of the wave amplitude even before the shock formation distance $x_{\mathrm{sh}}$ is observed. This amplitude decay is attributed to the numerical diffusion acting on the entire wave form as observed in Fig. \ref{fig:correctorWeight} for $\gamma=0.5$. As it can be seen in Figs. \ref{fig:dxStudy_redblueShift} and \ref{fig:dxdtStudy_shock}, the numerical diffusion decreases with increasing spatial resolution for $\gamma=0.5$. The same trend is observed in Fig. \ref{fig:longSimu_dxStudy} as well, where for high spatial resolution, only a very small amplitude decay is observed in the range $x<x_{\mathrm{sh}}$. However, even for the high spatial resolutions, a pronounced decay of the shock amplitude is observed in the range $x>x_{\mathrm{sh}}$. It is further observed that the envelope of the decaying amplitude converges with increasing spatial resolution. The underlying mechanism of this amplitude decay is related to the formation of the characteristic saw-tooth pattern and requires a more detailed explanation.

Physically, the attenuation of the wave amplitude is governed by viscous forces and thermal conduction \citep{Shevchenko_and_Kaltenbacher_2015}. \citet{Fay_1931} argued that, in many situations, the thermal conduction does not significantly contribute to the attenuation process and showed that the viscous loss is proportional to the curvature of the phase velocity profile of the wave. Therefore, the viscous loss is zero in regions of constant velocity gradient, and the attenuation in regions of nonzero curvature causes the wave to develop a profile of constant velocity gradient. In conjunction with the fact that the phase velocity increases with increasing pressure, this explains the formation of the characteristic saw-tooth pattern, involving a shock at the wave front \citep{Fay_1931}. The mechanism to explain the amplitude decay observed in Fig. \ref{fig:longSimu_dxStudy} is related to the stabilization of the shock front. The formation of the shock is associated with the transfer of wave energy to higher harmonics, where typically, the first ten harmonics are clearly noticeable in the wave amplitude spectrum at the shock formation distance \citep{Handbook_of_Acoustics, Treeby_et_al_2020}. Generally, the attenuation of sound waves can effectively be described by power laws of the wave frequencies, which essentially reflects the increase of viscous attenuation with increasing curvature of the wave profile. Thus, the higher harmonics of the wave are attenuated more strongly than the lower harmonics \citep{Chen_2005, Jimenez_et_al_2015, Liu_et_al_2018}. An idealized shock contains an infinite amount of harmonics. However, \citet{Fay_1931} showed that the increasing attenuation at the developing shock front imposes a limit on the energy transfer to higher harmonics in a way that prevents fluid particles in the shock front to overtake the foot of the wave, thereby stabilizing the shock front. Building up on this line of reasoning, \citet{Rudnick_1952} argued that if a stable shock front is present, then the rate of attenuation of such a stabilized shock must be independent of the process that leads to the presence of the stabilized shock front. In the present work, the numerical diffusion introduced by the predictor-corrector method is the stabilizing attenuation process. We argue that in the limit of a stabilized shock, this artificial numerical diffusion can mimic the actual physical attenuation process as it attenuates the harmonics at the same rate at which they feed into the wave front. In the remaining profile of the saw-tooth wave, the curvature and therefore the attenuation is approximately zero, so that the amplitude decay is exclusively governed by the attenuation of the higher harmonics feeding into the shock. In order to demonstrate that the shock attenuation observed in Fig. \ref{fig:longSimu_dxStudy} indeed follows theoretical predictions, we employ the following analytical solution, derived by \citet{Blackstock_1966} based on the work by \citet{Whitham_1952}, for the decay of the shock amplitude of a plane 1d wave in the limit of large $x/x_{\mathrm{sh}}$:
\begin{equation}
\dfrac{\phi \left(x,t\right)}{\Delta \phi_{\mathrm{a}}} = \dfrac{\pi}{1+\dfrac{x}{x_{\mathrm{sh}}}}.
\label{ampDecay}    
\end{equation}
In Eq. \eqref{ampDecay}, $\Delta \phi_{\mathrm{a}}$ is the excitation amplitude of a sinusoidal wave of the form $\phi = \Delta \phi_{\mathrm{a}} \sin\left(2\pi f_0 t\right)$. Taking into account that the continuous excitation function given by Eq. \eqref{contiExcitation} has an amplitude of $\Delta \phi_{\mathrm{a}}= \Delta p_{\mathrm{a}}/2$, the dashed red envelope in Fig. \ref{fig:longSimu_dxStudy} is obtained. The range of validity of the analytical reference solution is approximately given by $x> 3.5x_{\mathrm{sh}}$, where $3.5x_{\mathrm{sh}}$ is the approximate distance at which the originally sinusoidal wave has effectively degenerated into the saw-tooth shape \citep{Blackstock_1966}. It is observed that the shock amplitude indeed follows the envelope predicted by Eq. \eqref{ampDecay}. It is unclear why the wave trough is slightly raised relative to the envelope. Nevertheless, given that the shock front is represented on a grid of finite size, the agreement is encouraging.
\begin{figure}[htb]
\begin{center}
{\includegraphics[width=0.8\linewidth]{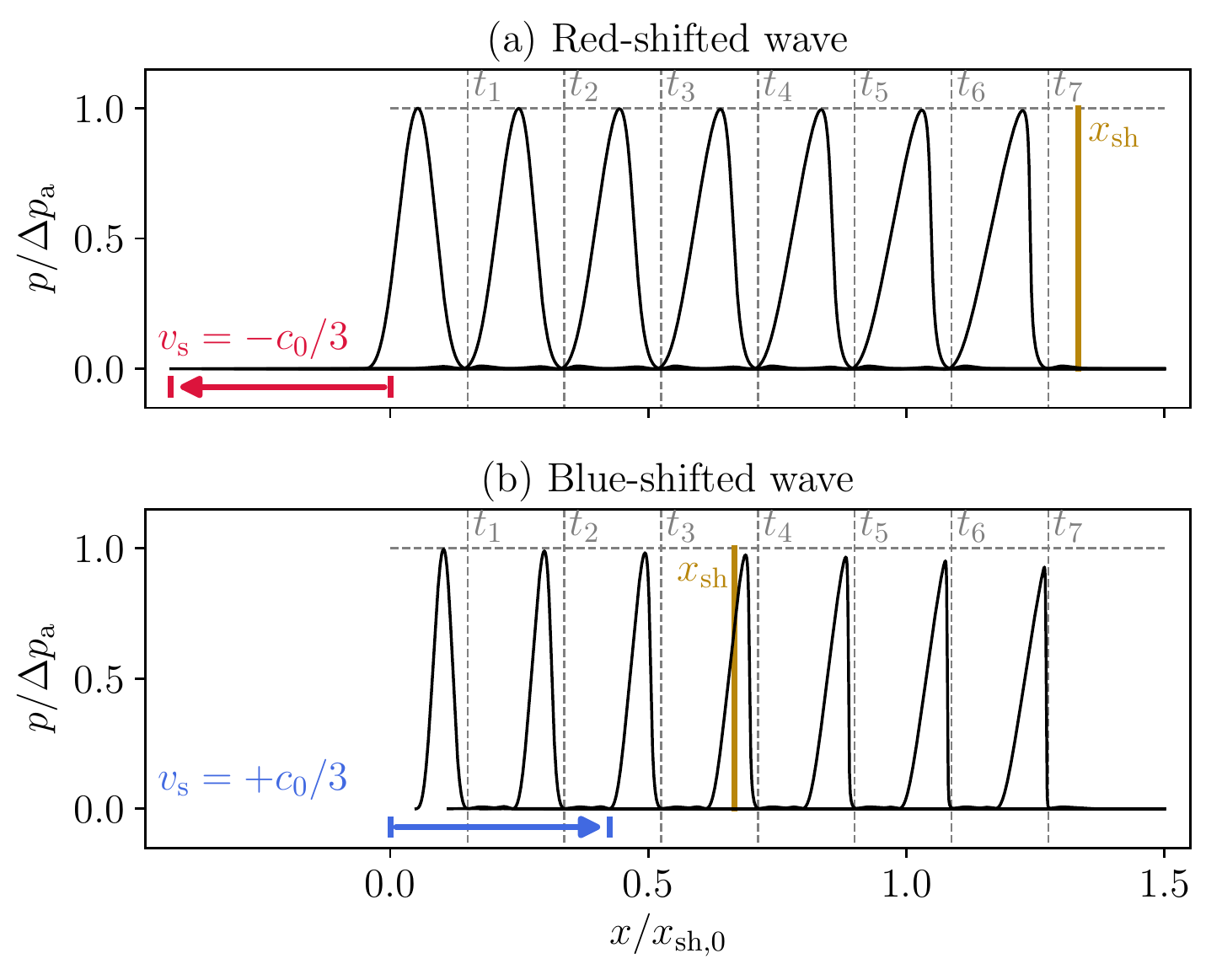}}
\caption{Pressure wave profiles obtained from the pulse excitation (Eq. \eqref{pulseExcitation}) for $\beta=9.0$ for the red-shifted wave (a) and the blue-shifted wave (b). The golden solid lines indicate the shock formation distances $x_{\mathrm{sh}}$ of the Doppler shifted waves predicted by Eq. \eqref{shockFormationDistance2}. The dashed vertical lines indicate the wave front positions based on $c_0$ at the time instants $t_1$ to $t_7$ at a spacing of $1.25/f_0$.}
\label{fig:shock_redblueShift}
\end{center}
\end{figure}

\begin{figure}[htb]
\begin{center}
{\includegraphics[width=0.8\linewidth]{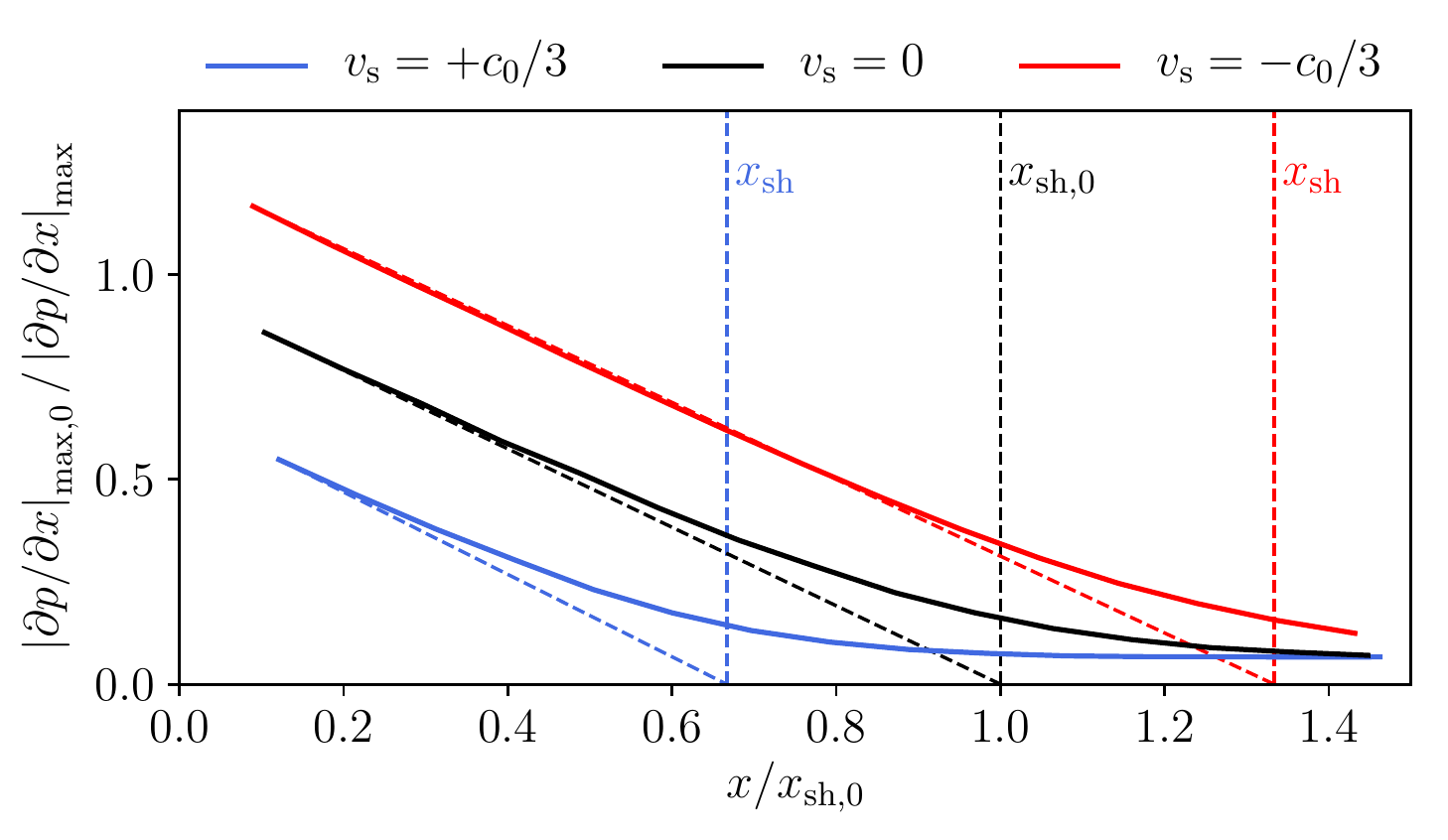}}
\caption{Evolution of the inverse of the maximum pressure gradient obtained from the pulse excitation (Eq. \eqref{pulseExcitation}) for the resting wave emitting boundary ($v_{\mathrm{s}}=0$), the red-shifted wave, and the blue-shifted for $\beta=9.0$. $\left|\partial p/\partial x\right|_{\mathrm{max},0}$ is the maximum gradient of the undistorted wave profile ($v_{\mathrm{s}}=0$, $\beta=0$). The vertical dashed lines indicate the shock formation distances $x_{\mathrm{sh}}$ of the Doppler shifted waves predicted by Eq. \eqref{shockFormationDistance2}. The transverse dashed lines connect the shock formation distances on the $x$-abscissa with the starting points of the corresponding inverse gradient slopes and form the theoretical asymptotic evolutions of the inverse gradient.}
\label{fig:invSlope}
\end{center}
\end{figure}

Fig. \ref{fig:shock_redblueShift} shows the instantaneous wave profiles at the same time instants as in Fig. \ref{fig:dxdtStudy_shock}, again obtained from the pulse excitation function given by Eq. \eqref{pulseExcitation}, however with the wave emitting boundary moving at constant speed $v_{\mathrm{s}} \pm c_0/3$. The corresponding shock formation distances predicted by Eq. \eqref{shockFormationDistance2} are $x_{\mathrm{sh}}/x_{\mathrm{sh},0} = \left\{2/3; \: 4/3\right\}$, respectively, and are indicated by the golden vertical lines in Fig. \ref{fig:shock_redblueShift}. It is observed that, as predicted, the blue-shifted wave profile (b) undergoes a more pronounced wave steepening than the red-shifted wave profile (a). In both cases, the wave amplitude remains approximately constant up to the shock formation distance. The travel distance of the blue-shifted wave exceeds the corresponding shock formation distance, and, as a result, the blue-shifted wave experiences an amplitude decay in the range $x > x_{\mathrm{sh}}$ due to the mechanism discussed above. Similar to the approach presented in the work by \citet{Chandrasekaran_2021}, we employ the maximum absolute value of the inverse of the wave slope, here denoted by the maximum inverse pressure gradient $\left(\left|\partial p/\partial x\right|_{\mathrm{max}}\right)^{-1}$, to indicate the Doppler shift of the shock formation distance more clearly. Fig. \ref{fig:invSlope} shows the evolution of $\left(\left|\partial p/\partial x\right|_{\mathrm{max}}\right)^{-1}$ over $x$ for the resting wave emitting boundary ($v_{\mathrm{s}}=0$), and for the blue and the red-shifted waves ($v_{\mathrm{s}}=\pm c_0/3$), respectively. Again, the pulse excitation function given by Eq. \eqref{pulseExcitation} is applied. As noted by \citet{Chandrasekaran_2021}, the inverse slope is expected to decay linearly over $x$, approaching 0 at the shock formation distance $x_{\mathrm{sh}}$. The linear decay behavior is clearly visible well before the shock formation distance is reached. For reference, the individual shock formation distances on the $x$-abscissa are connected to the corresponding starting points of the trajectories by the transverse dashed lines. The trajectories of the maximum inverse pressure gradient follow the theoretically linear slope up to a certain distance before the shock formation distance, where the trajectories flatten out gradually. Also, the trajectories converge to a value larger than zero, which is explained by the finite spatial resolution.

\subsection{Nonlinear Wave Propagation in 3D Spherical Symmetry}\label{sec:Nonlinear Wave Propagation in 3D Spherical Symmetry}

\begin{figure}[htb]
\begin{center}
{\includegraphics[width=0.75\linewidth]{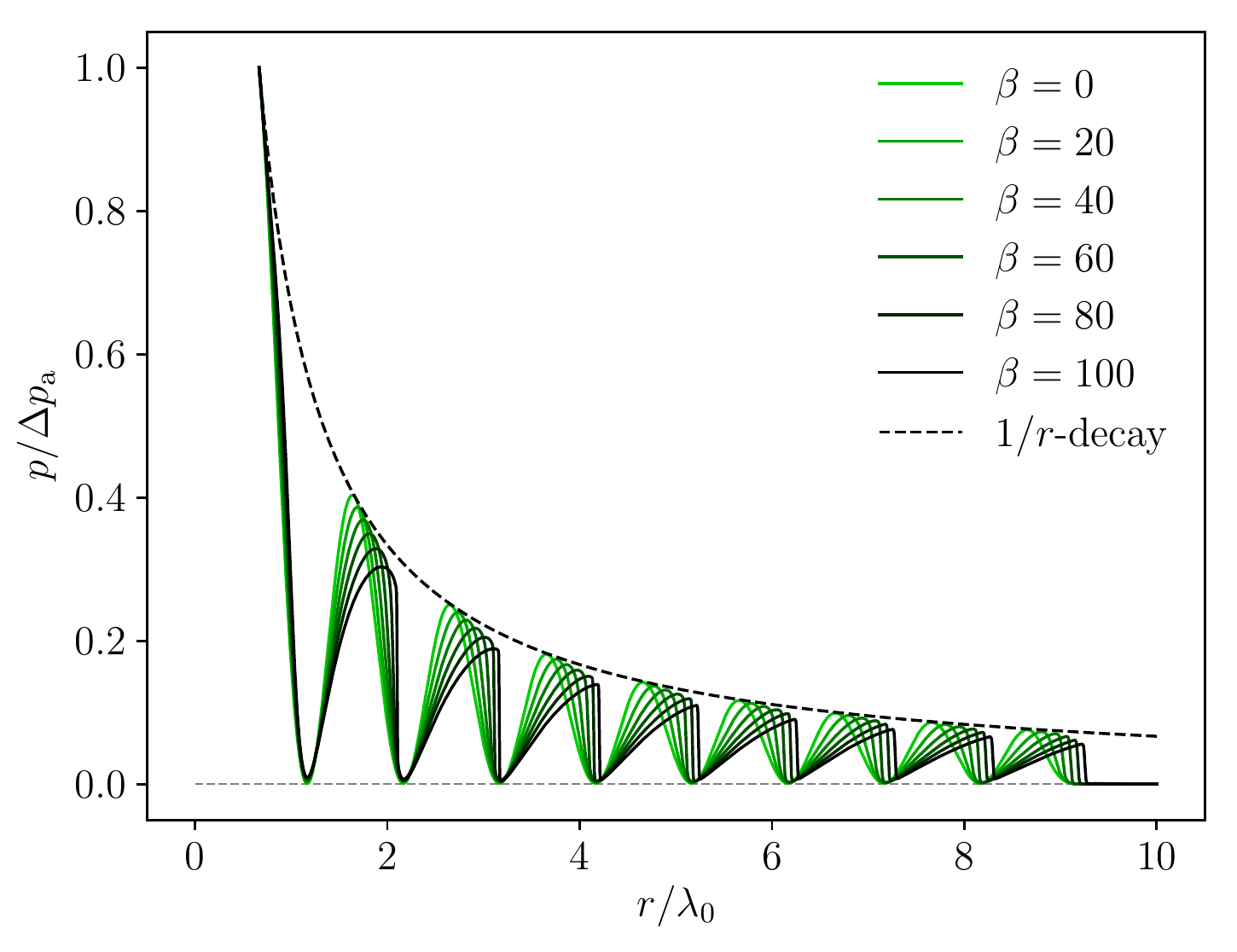}}
\caption{Instantaneous profile of the spherical wave obtained from continuous excitation (Eq. \eqref{contiExcitation}) at the fixed left boundary located at $R=2\lambda_0/3$. The nonlinearity coefficient $\beta$ is varied systematically. The black dashed line indicates the $1/r$-decay of a linear wave.}
\label{fig:betaStudySpherical}
\end{center}
\end{figure}

Fig. \ref{fig:betaStudySpherical} shows the spherical decay of a nonlinear wave for a systematic variation of the nonlinearity coefficient $\beta$. The wave is excited continuously (Eq. \eqref{contiExcitation}) at the fixed boundary ($\Delta v_{\mathrm{s,a}}=0$) at $R=2\lambda_0/3$. For $\beta=0$, the wave propagation is linear and the wave amplitude follows the $1/r$-decay law as indicated by the dashed line. The $1/r$-decay in 3d spherical symmetry corresponds to a propagation at constant amplitude in the Cartesian case. As $\beta$ increases, the wave profile is increasingly tilted in the direction of propagation and eventually forms a shock for large values of $\beta$. Compared to the 1d case, larger values of $\beta$ are needed in the 3d case for a shock to form. This is due to the geometrical amplitude decay, which weakens the nonlinear term of the Westervelt equation. Also, the decay of the shock amplitude relative to the $1/r$ slope is less pronounced than the shock amplitude decay in 1d, which is again explained by the increasingly linear behavior in the direction of outward propagation. 

\begin{figure}[htb]
\begin{center}
{\includegraphics[width=0.7\linewidth]{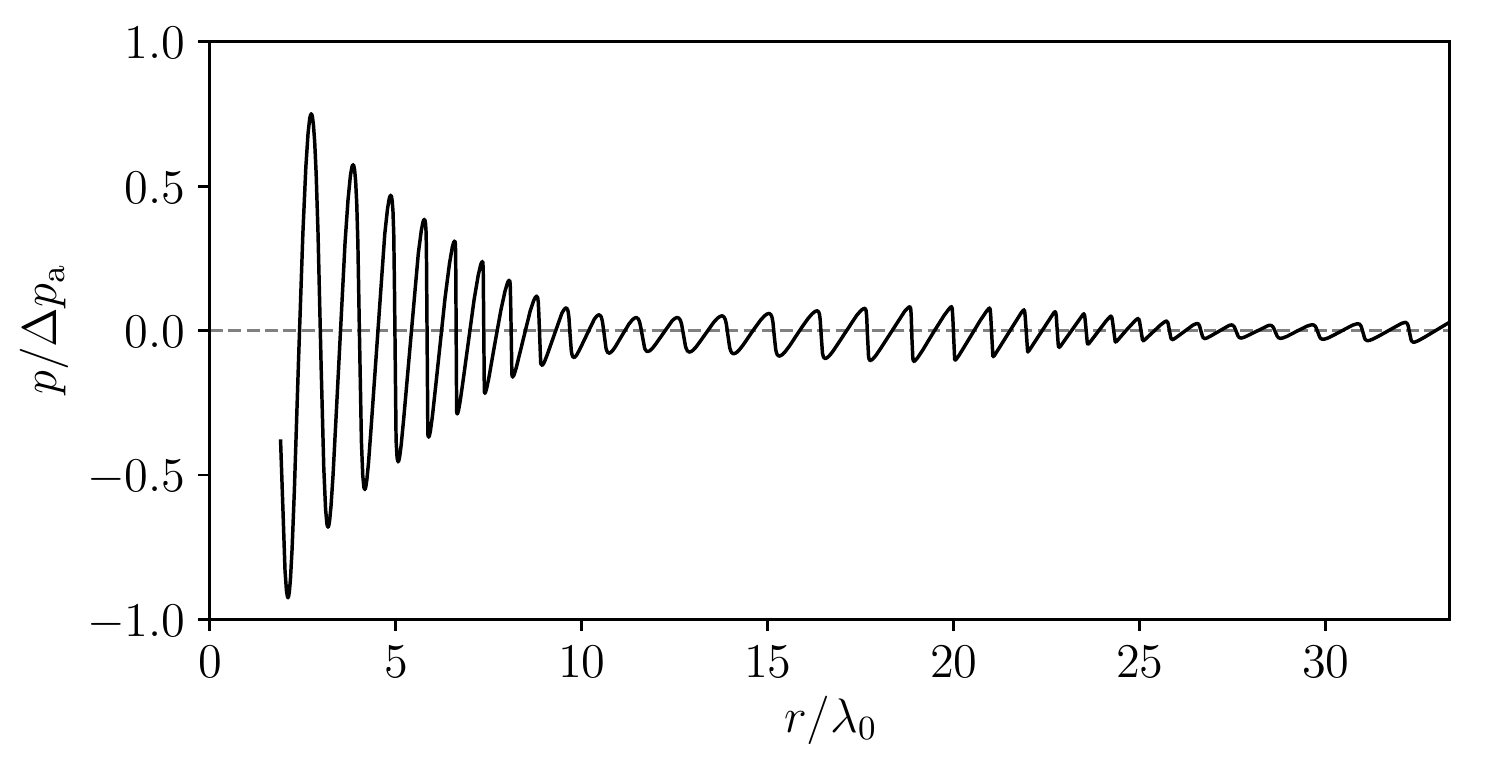}}
\caption{Instantaneous pressure wave profile of the spherical wave obtained from continuous excitation (Eq. \eqref{contiExcitation2}) at the moving boundary for $\beta=15$. The wave emitting boundary is in oscillatory motion, where $f_{\mathrm{s}}/f_0=3/50$ and $\Delta v_{\mathrm{s,a}}/c_0=4/15$. The initial radius $R_0 = X_0$ in Eq. \eqref{boundaryMotion} is chosen such that the peak negative deflection of the boundary is at $r=2\lambda_0/3$.}
\label{fig:pressureProfile_v400_spherical_shock_beta15}
\end{center}
\end{figure}

\begin{figure}[htb]
\begin{center}
{\includegraphics[width=0.7\linewidth]{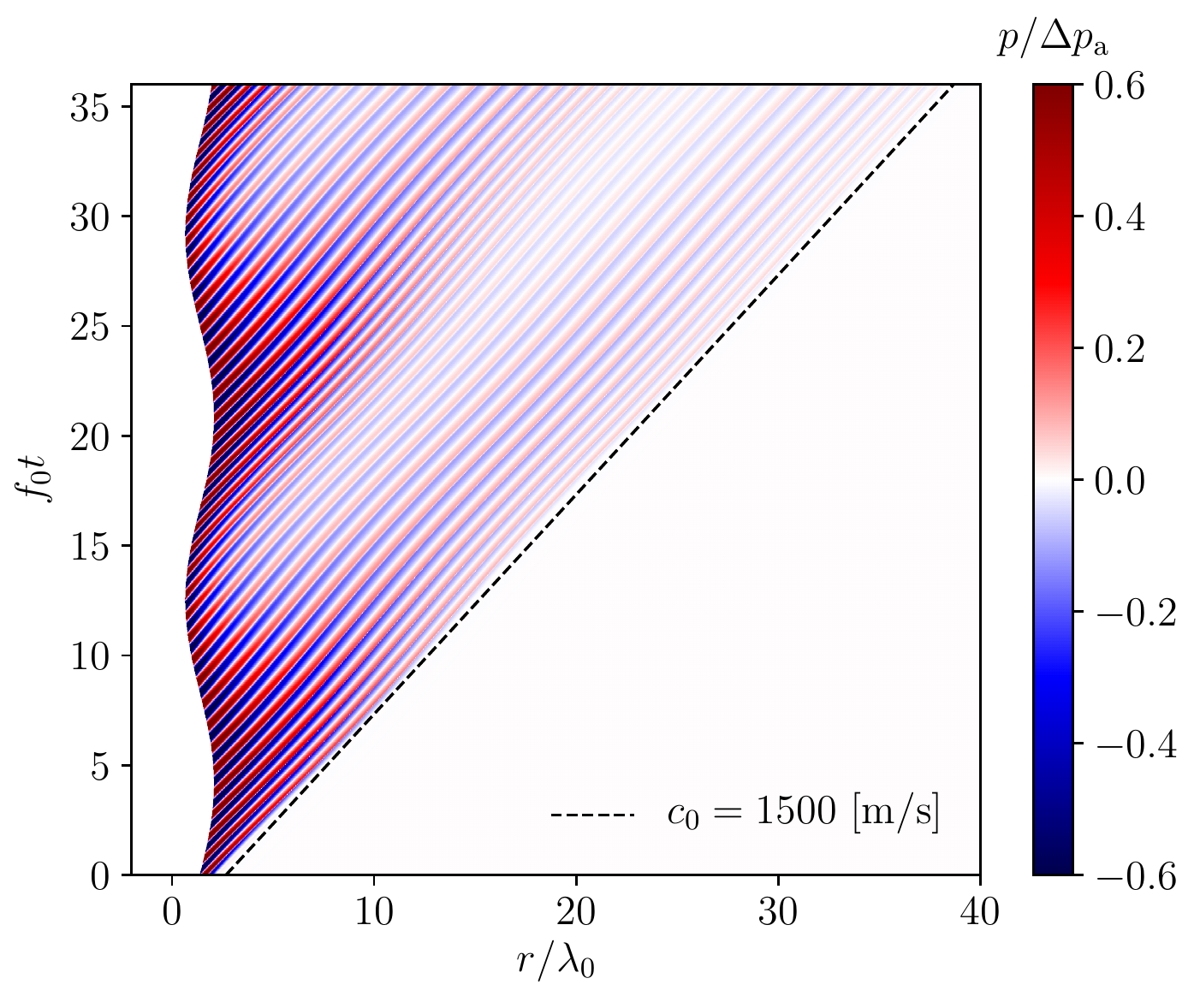}}
\caption{Pressure distribution in the $r$-$t$ plane obtained from continuous excitation (Eq. \eqref{contiExcitation2}) at the left boundary for $\beta=15$. The wave emitting boundary is in oscillatory motion around $R_0 = X_0 = 0$ according to Eq. \eqref{boundaryMotion}, where $f_{\mathrm{s}}/f_0=3/50$ and $\Delta v_{\mathrm{s,a}}/c_0=4/15$. The initial radius $R_0 = X_0$ in Eq. \eqref{boundaryMotion} is chosen such that the peak negative deflection of the boundary is at $r=2\lambda_0/3$. The dashed black line indicates the direction of propagation in the $x$-$t$ plane associated with a constant propagation speed of $c_0$.}
\label{fig:xtPlot_tremolo_v400_spherical_shock_beta15}
\end{center}
\end{figure}

Fig. \ref{fig:pressureProfile_v400_spherical_shock_beta15} shows an instantaneous profile of a spherical wave excited by an oscillating wave emitting boundary for $\beta=15$. The velocity amplitude of the emitting boundary is $\Delta v_{\mathrm{s,a}}/c_0=4/15$ with a corresponding frequency of $f_{\mathrm{s}}=\left(3/50\right)f_0$, and the initial radius $R_0 = X_0$ in Eq. \eqref{boundaryMotion} is chosen such that the peak negative deflection of the boundary is at $r=2\lambda_0/3$. Contrary to the previous 3d case, the wave amplitude does not decay strictly monotonically. Similar to the 1d case presented in Fig. \ref{fig:tremolo}, the excitation of a broad frequency band caused by the oscillatory boundary motion results in a breathing mode of the enveloping wave form. At a certain distance from the source, where the rate of amplitude decay due to the $1/r$-decay law has weakened sufficiently, the breathing mode causes the wave amplitude to increase and decrease in an oscillatory fashion, even in 3d. Also, the wave profile exhibits an oscillatory deformation between a rather sinusoidal pattern, associated with linear wave behavior, and a saw-tooth-like pattern, associated with nonlinear wave behavior. Again, this is explained by the oscillatory Doppler shift in the wavelength, which affects the rate of nonlinear wave steepening. In the absence of physical attenuation ($\delta =0$ in Eq. \eqref{Westervelt_eqn}), the transition from nonlinear to linear wave propagation is caused by the numerical attenuation of harmonics in the shock front by the diffusive predictor-corrector method. The corresponding pressure distribution in the $r$-$t$ plane is shown in Fig. \ref{fig:xtPlot_tremolo_v400_spherical_shock_beta15}. Close inspection again shows, similar to Fig. \ref{fig:xtPlot_tremolo_v0_v400}, a slight oscillation of the wave propagation speed relative to the constant propagation speed $c_0$ indicated by the black dashed line. Again, it is noted that this effect is predicted by \citet{Christov_2017}.

\section{Discussion}\label{sec:Discussion}

In the following, the feasibility of the presented methodology is discussed in the light of numerical accuracy and stability. Furthermore, the effects of nonlinear Doppler modulation of linear waves, linear Doppler modulation of nonlinear waves, and the combination of both are discussed.

\subsection{Numerical Stability vs. Accuracy}\label{sec:Numerical Stability vs. Accuracy}

A coordinate transformation, similar to the approach by \citet{Gasperini_et_al_2021}, allows to account for an accelerating wave emitting boundary in wave propagation. In the present study, the approach is extended to the nonlinear Westervelt equation in 1d domains and 3d spherical symmetry. The approach allows to simulate the propagation of nonlinear waves emitted from a moving boundary on moving computational grids. The transformed Westervelt equation (see Eqs. \eqref{LinEqn} to \eqref{kinCoupling}) is solved with an explicit FDTD method. The predictor corrector-method by \citet{Dey_and_Dey_1983} is employed to increase the numerical stability of the method, which is of particular importance when the formation of shocks and rapid grid motion/deformation is involved. \citet{Nascimento_et_al_2010} successfully applied the anti-dispersive predictor-corrector method by \citet{Dey_and_Dey_1983} to the linear wave equation subject to noisy seismic signals.

In the present work, the scheme is shown to successfully stabilize the explicit solution when sharp features, i.e., shocks, are present. In Fig. \ref{fig:longSimu_dxStudy}, the wave traveling distance involves approximately 12 shock formation distances without any signs of dispersive numerical noise. However, as demonstrated by Fig. \ref{fig:correctorWeight} for the linear wave, the enhanced numerical stability comes at the cost of numerical diffusion, which must be counteracted by an increase of spatial resolution. For 80 grid points per wavelength $\lambda_0$ and the recommended corrector weight of $\gamma=0.5$, the wave amplitude decreases by approximately 9\% of its starting value over a traveling distance of $7.5\lambda_0$. Based on the latter observation and the convergence studies for the nonlinear wave (see Fig. \ref{fig:dxdtStudy_shock}) and the Doppler-shifted linear wave (see Fig. \ref{fig:dxStudy_redblueShift}), a minimum of 160 grid points per initial wavelength is recommended to avoid excessive numerical diffusion when the predictor-corrector step is applied with a corrector weight of $\gamma=0.5$. For a linear wave, the recommended spatial resolution is significantly higher than the required 10 grid points per characteristic wavelength reported by \citet{Huijssen_and_Verwij_2010}, \citet{Wang_et_al_2012}, and \citet{Doinikov_et_al_2014} for finite difference methods. However, Fig. \ref{fig:longSimu_dxStudy} shows that with a resolution of 320 grid points per initial wavelength, the scheme is capable of accurately representing the rate of shock amplitude decay predicted by \citet{Whitham_1952} and \citet{Blackstock_1966}. 

The observed order of convergence is $\mathcal{O}\left(\Delta t\right)$ in $\Delta t$ and between $\mathcal{O}\left(\Delta \xi\right)$ and $\mathcal{O}\left(\Delta \xi^2\right)$ in $\Delta \xi$ (see Figs. \ref{fig:convDiagDoppler} and \ref{fig:convDiagShock}), suggesting a terminal convergence behaviour of second order in $\Delta \xi$. The temporal convergence rate, on the other hand, is of first order. Given the sixth order accurate finite difference approximation of the spatial derivatives, the finite difference approximation of the time derivatives in conjunction with the predictor-corrector method appears to hold the biggest potential for improvement. The discussion in Sec. \ref{sec:Predictor-Corrector Method} and in particular Eq. \eqref{LinEqn_antiDispersion} point towards two options to increase the order of convergence. The first option builds upon our present approach, in which, strictly following the predictor-corrector steps by \citet{Dey_and_Dey_1983} as represented by Eqs. \eqref{LinEqn_discr2} and \eqref{LinEqn_discr_corr}, the term $\gamma\mathcal{H}$ in Eq. \eqref{LinEqn_antiDispersion} is implicitly taken into account and a first order accurate finite difference approximation of the time derivatives is applied. \citet{Dey_1999} shows that for a corrector weight of $\gamma=0.5$, this particular configuration of the predictor-corrector method formally evolves into a second order Runge-Kutta method, however for ordinary differential equations of the form $\dot y=f\left(y\right)$ and partial differential equations of the form $\partial y/\partial t = f\left(y, \partial^k y/\partial x^k\right)$, such as Burger's equation. In the present work, we apply the method to a linearized partial differential equation involving higher order time derivatives and mixed spatial/temporal derivatives due the coordinate transformation on physical space (see the transformed Westervelt equation given by Eq. \eqref{LinEqn}), which likely explains the loss of second order accuracy for $\gamma=0.5$. In this regard, a higher order of convergence could possibly be achieved by an improved treatment of the mixed derivatives, which are treated as source terms in the present approach, or by a higher order Taylor expansion in Eq. \eqref{nonLinTerm}. The second option builds upon the approach by \citet{Nascimento_et_al_2010}. Translated to the present work, \citet{Nascimento_et_al_2010} omit the term $\gamma\mathcal{H}$ in Eq. \eqref{LinEqn_antiDispersion} and only keep the additional damping term given by the term in square brackets in Eq. \eqref{LinEqn_antiDispersion}. This modification of the method by \citet{Dey_and_Dey_1983} is not associated with a Runge-Kutta-type time integration, and any arbitrary order of accuracy can be chosen for the finite difference approximation of the time derivatives. However, additional spectral optimization of the finite difference scheme \citep{Tam_and_Webb_1993} might be necessary in order to suppress additional numerical noise if the order of accuracy of the temporal finite differences is increased.

\citet{Fay_1931} and \citet{Rudnick_1952} showed that the formation and stabilization of a saw-tooth wave requires the attenuation of the higher harmonics that feed into the shock front. Keeping in mind that the viscous attenuation acting on the ramp of the saw-tooth wave is zero due to the zero curvature \citep{Fay_1931}, the attenuation to stabilize the waveform must exclusively take place in the shock front. Furthermore, the attenuation rate in the shock front must balance the rate at which energy feeds into the shock, which again is governed by the particle velocity distribution of the wave profile. Hence, the presence of the stable shock front itself dictates its amplitude decay rate, irrespective of the mechanism that is responsible for the stabilization of the shock \citep{Rudnick_1952}. Physically, the viscous attenuation of the higher wave harmonics is the main contributor to the shock stabilization \citep{Fay_1931}. However, the shock stabilization can also be achieved by the numerical diffusion imposed by the predictor-corrector method by \citet{Dey_and_Dey_1983} or any other suitable numerical scheme. Therefore, we argue that in the limit of strongly nonlinear wave propagation, which manifests in the formation of a stabilized shock, the numerical diffusion can correctly mimic the attenuation by physical viscosity, provided that the spatial resolution is high enough so that excessive numerical diffusion of the lower harmonics associated with the continuous part of the wave profile is avoided. In summary, the predictor-corrector method by \citep{Dey_and_Dey_1983} is found to be advantageous in the context of nonlinear acoustic wave propagation, because the associated numerical diffusion decreases with increasing spatial resolution for a fixed $\mathrm{CFL}$ number, while the scheme still maintains its capability to stabilize shocks. In that sense, it acts as a low-pass filter, where the cut-off frequency increases with increasing spatial resolution. This makes the approach in the present work attractive for situations that involve strong nonlinear propagation behavior and/or wave emitting boundary motion, where any method would require a higher spatial resolution in order to represent the higher wave harmonics. Concerning the weights of the predicted and the corrected solutions in Eq. \eqref{LinEqn_discr_corr}, a value of $\gamma=0.5$ is recommended as discussed in Sec. \ref{sec:Linear Doppler Modulation of the Linear Wave}.

The switch between 1d and 3d spherical symmetry is straightforward in the present framework. The 1d equations are complemented by the additional term $\mathcal{A}_{\mathrm{G},r}$ given by Eq. \eqref{coeffAr} to obtain the 3d spherically symmetric representation. The nonlinear contributions as well as the kinematic coupling terms resulting from the coordinate transformation remain unaffected.

\subsection{Nonlinear Wave Propagation}\label{sec:Nonlinear Wave Propagation}

The nonlinear Doppler modulation of the linear wave ($\beta=0$) results in a nonlinear frequency and wave amplitude modulation, loosely referred to as a combined vibrato and tremolo by \citet{Christov_2017}. For the given ratio $f_{\mathrm{s}}/f_0=3/50$ of source motion and sound wave frequency, and for a velocity amplitude $\Delta v_{\mathrm{s,a}}$ that is small relative to the sound speed $c_0$, the waveform of the amplitude modulation in Fig. \ref{fig:tremolo} is in good agreement with the spatial distribution of the asymptotic amplitude modulation factor $\mathrm{AM}\left(x,t\right)$ derived by \citet{Christov_2017} for small wave emitting boundary motion. It can be seen from Eq. \eqref{AM} that at a fixed time instant, the spatial waveform of the amplitude modulation $\mathrm{AM}\left(x,t\right)$ contains only one harmonic. However, Eq. \eqref{AM} is only valid for small velocity amplitudes of the moving wave emitting boundary \citep{Christov_2017}. With increasing velocity amplitude $\Delta v_{\mathrm{s,a}}$, additional frequencies are excited and the spatial wave profile develops a broadband spectrum of increasing bandwidth, which is demonstrated by the pressure amplitude spectrum presented in Fig. \ref{fig:frequ_v0_v200_v400}. The presence of a broadband spectrum for larger wave emitting boundary motion is in line with the finding by \citet{Gasperini_et_al_2021}. Despite the linearity of the medium ($\beta=0$) for this test case, the pressure distribution in the $x-t$-plane in Fig. \ref{fig:xtPlot_tremolo_v0_v400} indicates small oscillations of the wave propagation speed around $c_0$. This effect is also predicted and demonstrated by \citet{Christov_2017}. Fig. \ref{fig:tremolo} further shows that the amplitude modulation predicted by the asymptotic theory provided by \citet{Christov_2017} increases with increasing velocity amplitude $\Delta v_{\mathrm{s,a}}$, and that the excitation of the broad frequency band due to larger boundary motion generates amplitudes larger than those predicted by the asymptotic theory for small motion. Fig. \ref{fig:pressureProfile_v400_spherical_shock_beta15} indicates that this can lead to an increase in amplitude even in 3d spherical symmetry. This is explained by the fact that for the $1/r$-decay law, the change of geometrical attenuation over $r$ is proportional to $dr^{-1}/dr = -r^{-2}$. Hence, it decreases over $r$, so that the geometrical attenuation is ultimately obscured by the amplitude modulation caused by the oscillation of the wave emitting boundary.

The effect of a wave emitting boundary moving at constant speed on the nonlinear steepening of the wave for $\beta>0$ is summarized in Fig. \ref{fig:invSlope}. The figure shows that the inverse of the maximum pressure gradient of the wave profile decays linearly over the wave traveling distance, approaching the shock formation distance on the $x$-abscissa, which is in line with the analysis by \citet{Chandrasekaran_2021}. The finite mesh resolution is likely to explain why the slope gradually flattens out in close vicinity to the shock formation distance. Most importantly, Fig. \ref{fig:invSlope} demonstrates that the rate of nonlinear wave steepening experiences a linear Doppler shift, when the wave emitting boundary moves at constant velocity through the resting background medium. The linear Doppler shift of nonlinear wave steepening is manifest in the parallel shift of the corresponding inverse pressure gradient slopes. The linear Doppler shift of the shock formation distance is predicted analytically by correcting the standard formula by the Doppler shift in the wave frequency (see Eq. \eqref{shockFormationDistance2}). The wave profile excited by the oscillating spherical boundary in Fig. \ref{fig:pressureProfile_v400_spherical_shock_beta15} exhibits an oscillatory change between nonlinear wave steepening and wave flattening. The combined effects of this oscillatory deformation and the aforementioned amplitude oscillation can be seen as the nonlinear Doppler modulation of a finite amplitude wave propagating in a nonlinear medium.

\subsection{Prospective Future Extensions}\label{sec:Prospective Future Extensions}

The presented methodology can potentially be extended to perform detailed analysis of the propagation of pressure waves radiated from oscillating bubbles. However, this requires to take the fluid motion into account that is induced by the moving bubble interface. The motion of the bubble wall accelerates the surrounding liquid, so that the wave propagates through a background medium of non-stationary and non-uniform velocity. For low Mach numbers, the local plane wave (LPW) approximation \citep{Willatzen_2001} or composite solutions \citep{Androsov_et_al_2013} can be employed to account for the background motion. \citet{Godin_2011} provides an exact form of the linear wave equation for moving, non-stationary background media, and gives an overview of specializations thereof, which can potentially be incorporated into the existing framework. Together with the presented methodology for nonlinear wave propagation from a moving wave emitting boundary, this could provide more insights into the complicated behavior of nonlinear waves emitted from oscillating bubbles in a liquid.

Furthermore, the flexibility provided by the mapping function to achieve the transformation between the physical and the computational domain can be further exploited. For instance, the grid resolution can be adapted to sharp features of the solution \citep{Sewerin_Rigopoulos_2017, Huang_2011}, which might help to reduce the computational cost of the presented method.

\section{Conclusion}\label{sec:Conclusion}
By introducing a coordinate transformation, as presented by \citet{Gasperini_et_al_2021} for the linear wave equation, in the nonlinear lossless Westervelt equation, a generic equation is derived to describe the propagation of pressure waves generated by an accelerating wave emitting boundary in a nonlinear medium in 1d and 3d spherical symmetry. In the present study, the predictor-corrector method by \citet{Dey_and_Dey_1983} proved to be useful to achieve numerical stability in the presence of shocks and rapid grid motion, while still allowing for the correct asymptotic decay behavior of a propagating shock. However, the increased numerical stability comes at the cost of numerical diffusion that must be counteracted by an increase of spatial resolution. Therefore, the scheme is recommended for situations that involve strong nonlinear wave propagation behavior, either due to accelerating motion of the wave emitting boundary or due to the nonlinearity of the medium. It is shown that the nonlinear wave steepening, and therefore also the shock formation distance in 1d, is subject to a linear Doppler shift when the wave emitting boundary moves at constant speed, which is in line with the analytical prediction based on an adjusted formula for the shock formation distance. The nonlinear Doppler modulation of the wave frequency and amplitude caused by a spatially oscillating wave emitting boundary is shown to be more complicated. For small motion, the amplitude modulation is in good agreement with the asymptotic theory by \citet{Christov_2017}. For larger velocity amplitudes of the sound radiating source, the resulting wave profile is shown to develop a broadband amplitude spectrum, even when the radiated wave is purely sinusoidal and when the medium is linear. This is in line with the findings of \citet{Gasperini_et_al_2021} and indicates that more research on the complicated behavior of nonlinear pressure waves emitted or scattered from accelerating sources is needed. For instance, the excitation of the broadband spectrum caused by the oscillating wave emitting boundary likely has an effect on the attenuation of the pressure wave, which is frequency dependent \citep{Chen_2005, Liu_et_al_2018, Jimenez_et_al_2015}. In order to achieve a more accurate representation of nonlinear waves radiated by oscillating bubbles, the inclusion of transsonic and supersonic background flows, induced by the bubble motion, into the existing modeling framework is one of the most pressing problems.

\section*{Acknowledgments}
This research was funded by the Deutsche Forschungsgemeinschaft (DFG, German Research Foundation), grant numbers 441063377 and 443546539.

\bibliographystyle{apalike}

\begin{thebibliography}{}

\bibitem[Amara et~al., 2013]{Amara_et_al_2013}
Amara, L., Berreksi, A., and Achour, B. (2013).
\newblock Adapted {MacCormack} finite-differences scheme for water hammer
  simulation.
\newblock {\em Journal of Civil Engineering and Science}, Vol.2:226--233.

\bibitem[Androsov et~al., 2013]{Androsov_et_al_2013}
Androsov, A., Harig, S., Fuchs, A., Immerz, A., Rakowsky, N., Hiller, W., and
  Danilov, S. (2013).
\newblock {\em Tsunami Wave Propagation}, pages 43--72.
\newblock IntechOpen.

\bibitem[Bailey et~al., 2003]{Bailey_et_al_2003}
Bailey, M., Khokhlova, V., Sapozhnikov, O., Kargl, S., and Crum, L. (2003).
\newblock Physical mechanisms of the therapeutic effect of ultrasound (a
  review).
\newblock {\em Acoust Phys}, 49:369--388.

\bibitem[Bernard, 1992]{Bernard_1992}
Bernard, R.~S. (1992).
\newblock A {MacCormack} scheme for incompressible flow.
\newblock {\em Computers \& Mathematics with Applications}, 24(5):151--168.

\bibitem[Blackstock, 1966]{Blackstock_1966}
Blackstock, D.~T. (1966).
\newblock Connection between the {Fay} and {Fubini} solutions for plane sound
  waves of finite amplitude.
\newblock {\em The Journal of the Acoustical Society of America},
  39(6):1019--1026.

\bibitem[Blackstock et~al., 1998]{Blackstock_et_al_1998}
Blackstock, D.~T., Hamilton, M.~F., and Pierce, A.~D. (1998).
\newblock Progressive waves in lossless and lossy fluids.
\newblock In Hamilton, M.~F. and Blackstock, D.~T., editors, {\em Nonlinear
  Acoustics}. Academic Press, San Diego.

\bibitem[Bouche et~al., 2003]{Bouche_2003}
Bouche, D., Bonnaud, G., and Ramos, D. (2003).
\newblock Comparison of numerical schemes for solving the advection equation.
\newblock {\em Applied Mathematics Letters}, 16(2):147--154.

\bibitem[Breuß, 2004]{Breuss_2004}
Breuß, M. (2004).
\newblock The correct use of the {Lax–Friedrichs} method.
\newblock {\em ESAIM: Mathematical Modelling and Numerical Analysis}, 38.

\bibitem[Chandrasekaran et~al., 2021]{Chandrasekaran_2021}
Chandrasekaran, N., Mercier, B., and Colonna, P. (2021).
\newblock {\em Formation of Rarefaction Shockwaves in Non-ideal Gases with
  Temperature Gradients}, volume~28, pages 20--25.
\newblock Springer, Cham.

\bibitem[Chen, 2005]{Chen_2005}
Chen, W. (2005).
\newblock Lévy stable distribution and [0,2] power law dependence of acoustic
  absorption on frequency in various lossy media.
\newblock {\em Chinese Physics Letters}, 22:2601.

\bibitem[Christov and Christov, 2017]{Christov_2017}
Christov, I.~C. and Christov, C.~I. (2017).
\newblock On mechanical waves and doppler shifts from moving boundaries.
\newblock {\em Mathematical Methods in the Applied Sciences},
  40(12):4481–4492.

\bibitem[Dey, 1999]{Dey_1999}
Dey, S. (1999).
\newblock A novel explicit finite difference scheme for partial differential
  equations.
\newblock {\em Mathematical Modelling and Analysis}, 4(1):70--78.

\bibitem[Dey and Dey, 1983]{Dey_and_Dey_1983}
Dey, S. and Dey, C. (1983).
\newblock An explicit predictor-corrector solver with applications to
  {Burgers}' equation: {NASA} technical memorandum 84402.
\newblock Technical report, National Aeronautics and Space Administration.

\bibitem[Doinikov et~al., 2014]{Doinikov_et_al_2014}
Doinikov, A.~A., Novell, A., Calmon, P., and Bouakaz, A. (2014).
\newblock Simulations and measurements of 3-d ultrasonic fields radiated by
  phased-array transducers using the {Westervelt} equation.
\newblock {\em IEEE Transactions on Ultrasonics, Ferroelectrics, and Frequency
  Control}, 61(9):1470--1477.

\bibitem[Dubey, 2016]{Dubey_2016}
Dubey, R.~K. (2016).
\newblock Data dependent stability of forward in time and centred in space
  ({FTCS}) scheme for scalar hyperbolic.
\newblock {\em International Journal of Numerical Analysis \& Modeling},
  13:689–704.

\bibitem[Fay, 1931]{Fay_1931}
Fay, R.~D. (1931).
\newblock Plane sound waves of finite amplitude.
\newblock {\em The Journal of the Acoustical Society of America},
  3(2A):222--241.

\bibitem[Fornberg, 1988]{Fornberg_1988}
Fornberg, B. (1988).
\newblock Generation of finite difference formulas on arbitrarily spaced grids.
\newblock {\em Mathematics of Computation}, 51(184):699--706.

\bibitem[Gasperini et~al., 2021]{Gasperini_et_al_2021}
Gasperini, D., Beise, H.-P.~P., Schroeder, U., Antoine, X., and Geuzaine, C.
  (2021).
\newblock A frequency domain method for scattering problems with moving
  boundaries.
\newblock {\em {Wave Motion}}, 102:102717.

\bibitem[Godin, 2011]{Godin_2011}
Godin, O. (2011).
\newblock An exact wave equation for sound in inhomogeneous, moving, and
  non-stationary fluids.
\newblock {\em OCEANS'11 - MTS/IEEE Kona, Program Book}.

\bibitem[Haigh et~al., 2012]{Haigh_et_al_2012}
Haigh, A., Treeby, B., and McCreath, E. (2012).
\newblock Ultrasound simulation on the cell broadband engine using the
  westervelt equation.
\newblock In {\em Algorithms and Architectures for Parallel Processing}, pages
  241--252.

\bibitem[Hallaj and Cleveland, 1999]{Hallaj_and_Cleveland_1999}
Hallaj, I.~M. and Cleveland, R.~O. (1999).
\newblock {FDTD} simulation of finite-amplitude pressure and temperature fields
  for biomedical ultrasound.
\newblock {\em The Journal of the Acoustical Society of America}, 105:L7--12.

\bibitem[Hamilton and Blackstock, 1988]{Hamilton_and_Blackstock_1988}
Hamilton, M.~F. and Blackstock, D.~T. (1988).
\newblock On the coefficient of nonlinearity $\beta$ in nonlinear acoustics.
\newblock {\em The Journal of the Acoustical Society of America}, 83(1):74--77.

\bibitem[Hendee and Ritenour, 2002]{Hendee_and_Ritenour_2002}
Hendee, W. and Ritenour, E. (2002).
\newblock {\em Ultrasound Waves}.
\newblock John Wiley \& Sons, Ltd.

\bibitem[Hixon, 1997]{Hixon_1997}
Hixon, R. (1997).
\newblock On increasing the accuracy of {MacCormack} schemes for aeroacoustic
  applications.
\newblock In {\em 3rd AIAA/CEAS Aeroacoustics Conference}.

\bibitem[Huang and Russell, 2011]{Huang_2011}
Huang, W. and Russell, R.~D. (2011).
\newblock {\em Adaptive Moving Mesh Methods}.
\newblock Springer.

\bibitem[Huijssen, 2008]{Huijssen_phd_2008}
Huijssen, J. (2008).
\newblock {\em Modeling of Nonlinear Medical Diagnostic Ultrasound}.
\newblock PhD thesis, Delft University of Technology.

\bibitem[Huijssen and Verweij, 2010]{Huijssen_and_Verwij_2010}
Huijssen, K. and Verweij, M. (2010).
\newblock An iterative method for the computation of nonlinear, wide-angle,
  pulsed acoustic fields of medical diagnostic transducers.
\newblock {\em The Journal of the Acoustical Society of America}, 127:33--44.

\bibitem[Jaros et~al., 2014]{Jaros_et_al_2014}
Jaros, J., Rendell, A., and Treeby, B. (2014).
\newblock Full-wave nonlinear ultrasound simulation on distributed clusters
  with applications in high-intensity focused ultrasound.
\newblock {\em International Journal of High Performance Computing
  Applications}, 30.

\bibitem[Jiménez et~al., 2015]{Jimenez_et_al_2015}
Jiménez, N., Redondo, J., Sánchez-Morcillo, V., and Camarena, F. (2015).
\newblock Nonlinear ultrasound simulations including complex frequency
  dependent attenuation.
\newblock {\em Physics Procedia}, 63:108--113.
\newblock 43rd Annual UIA Symposium 23—25 April 2014 CSIC Madrid, Spain.

\bibitem[Jing et~al., 2011]{Jing_et_al_2011}
Jing, Y., Shen, D., and Clement, G. (2011).
\newblock Verification of the {Westervelt} equation for focused transducers.
\newblock {\em Ultrasonics, Ferroelectrics and Frequency Control, IEEE
  Transactions on}, 58:1097 -- 1101.

\bibitem[Karamalis et~al., 2010]{Karamalis_et_al_2010}
Karamalis, A., Wein, W., and Navab, N. (2010).
\newblock Fast ultrasound image simulation using the {Westervelt} equation.
\newblock In Jiang, T., Navab, N., Pluim, J. P.~W., and Viergever, M.~A.,
  editors, {\em Medical Image Computing and Computer-Assisted Intervention --
  MICCAI 2010}, pages 243--250, Berlin, Heidelberg. Springer Berlin Heidelberg.

\bibitem[Kyriakou et~al., 2015]{Kyriakou_et_al_2015}
Kyriakou, A., Neufeld, E., Werner, B., Székely, G., and Kuster, N. (2015).
\newblock Full-wave acoustic and thermal modeling of transcranial ultrasound
  propagation and investigation of skull-induced aberration correction
  techniques: A feasibility study.
\newblock {\em Journal of Therapeutic Ultrasound}, 3:11.

\bibitem[Lauterborn et~al., 2014]{Handbook_of_Acoustics}
Lauterborn, W., Kurz, T., and Akhatov, I. (2014).
\newblock {\em Handbook of Acoustics}.
\newblock Springer Handbooks.

\bibitem[Lax and Wendroff, 1960]{Lax_Wendroff_1960}
Lax, P.~D. and Wendroff, B. (1960).
\newblock Systems of conservation laws.
\newblock {\em Communications on Pure and Applied Mathematics}, 13:217--237.

\bibitem[Liseikin, 2017]{Liseikin_2017}
Liseikin, V. (2017).
\newblock {\em Grid Generation Methods}.
\newblock Springer.

\bibitem[Liu et~al., 2009]{Liu_et_al_2009}
Liu, F., Zhang, G., Morton, S.~A., and Leveille, J.~P. (2009).
\newblock An optimized wave equation for seismic modeling and reverse time
  migration.
\newblock {\em GEOPHYSICS}, 74(6):WCA153--WCA158.

\bibitem[Liu et~al., 2018]{Liu_et_al_2018}
Liu, S., Yang, Y., Li, C., Guo, X., Tu, J., and Zhang, D. (2018).
\newblock Prediction of hifu propagation in a dispersive medium via
  khokhlov–zabolotskaya–kuznetsov model combined with a fractional order
  derivative.
\newblock {\em Applied Sciences}, 8:609.

\bibitem[MacCormack, 1982]{MacCormack_1982}
MacCormack, R.~W. (1982).
\newblock A numerical method for solving the equations of compressible viscous
  flow.
\newblock {\em AIAA Journal}, 20:1275–1281.

\bibitem[Machalińska-Murawska and Szydłowski, 2013]{Murawska_et_al_2013}
Machalińska-Murawska, J. and Szydłowski, M. (2013).
\newblock {Lax-Wendroff} and {McCormack} schemes for numerical simulation of
  unsteady gradually and rapidly varied open channel flow.
\newblock {\em Archives of Hydroengineering and Environmental Mechanics},
  60:51--62.

\bibitem[Meesala et~al., 2020]{Meesala_et_al_2020}
Meesala, V.~C., Hajj, M.~R., and Shahab, S. (2020).
\newblock Analysis and prediction of shock formation in acoustic energy
  transfer systems.
\newblock {\em Journal of Applied Physics}, 128(23):234902.

\bibitem[Miller et~al., 2012]{Miller_et_al_2012}
Miller, D., Smith, N., Bailey, M., Czarnota, G., Hynynen, K., and Makin, I.
  (2012).
\newblock Overview of therapeutic ultrasound applications and safety
  considerations.
\newblock {\em Journal of ultrasound in medicine : official journal of the
  American Institute of Ultrasound in Medicine}, 31:623--34.

\bibitem[Muir and Carstensen, 1980]{Muir_and_Carstensen_1980}
Muir, T. and Carstensen, E. (1980).
\newblock Prediction of nonlinear acoustic effects at biomedical frequencies
  and intensities.
\newblock {\em Ultrasound in Medicine \& Biology}, 6(4):345--357.

\bibitem[Na et~al., 2016]{Na_et_al_2016}
Na, F., Zhao, L., Xie, X.-B., Ge, Z., and Yao, Z.-X. (2016).
\newblock Two-dimensional time-domain finite-difference modeling for
  viscoelastic seismic wave propagation.
\newblock {\em Geophysical Journal International}, 206:1539--1551.

\bibitem[Nascimento and Pestana, 2010]{Nascimento_et_al_2010}
Nascimento, W. C.~R. and Pestana, R.~C. (2010).
\newblock An anti‐dispersion wave equation based on the predictor‐corrector
  method for seismic modeling and reverse time migration.
\newblock In {\em SEG Technical Program Expanded Abstracts}, pages 3226--3230.

\bibitem[Norton and Purrington, 2009]{Norton_and_Purrington_2009}
Norton, G.~V. and Purrington, R.~D. (2009).
\newblock The westervelt equation with viscous attenuation versus a causal
  propagation operator: A numerical comparison.
\newblock {\em Journal of Sound and Vibration}, 327(1):163--172.

\bibitem[Purrington and Norton, 2012]{Purrington_and_Norton_2012}
Purrington, R.~D. and Norton, G.~V. (2012).
\newblock A numerical comparison of the {Westervelt} equation with viscous
  attenuation and a causal propagation operator.
\newblock {\em Mathematics and Computers in Simulation}, 82(7):1287--1297.
\newblock Nonlinear Waves: Computation and Theory-X, WAVES 2009.

\bibitem[Qiao et~al., 2016]{Qiao_et_al_2016}
Qiao, S., Jackson, E., Coussios, C., and Cleveland, R. (2016).
\newblock Simulation of nonlinear propagation of biomedical ultrasound using
  {PZFLEX} and the {Khokhlov-Zabolotskaya-Kuznetsov} texas code.
\newblock {\em The Journal of the Acoustical Society of America},
  140:2039--2046.

\bibitem[Ramos and Nava, 2012]{Ramos_and_Nava_2012}
Ramos, J.~I. and Nava, E. (2012).
\newblock Numerical solution of the {Lighthill-Westervelt} equation. part 1: 1d
  problems.
\newblock In {\em Sociedade Portuguesa de Acústica}.

\bibitem[Rudnick, 1952]{Rudnick_1952}
Rudnick, I. (1952).
\newblock Theory of the attenuation of very high amplitude sound waves.
\newblock Technical report, Soundrive Engine Co.

\bibitem[Sewerin and Rigopoulos, 2017]{Sewerin_Rigopoulos_2017}
Sewerin, F. and Rigopoulos, S. (2017).
\newblock An explicit adaptive grid approach for the numerical solution of the
  population balance equation.
\newblock {\em Chemical Engineering Science}, 168:250--270.

\bibitem[Shampine, 2005]{Shampine_2005}
Shampine, L. (2005).
\newblock Two-step {Lax–Friedrichs} method.
\newblock {\em Applied Mathematics Letters}, 18(10):1134--1136.

\bibitem[Shevchenko and Kaltenbacher, 2015]{Shevchenko_and_Kaltenbacher_2015}
Shevchenko, I. and Kaltenbacher, B. (2015).
\newblock Absorbing boundary conditions for nonlinear acoustics: The westervelt
  equation.
\newblock {\em J. Comput. Phys.}, 302:200--221.

\bibitem[Solovchuk et~al., 2013]{Solovchuk_et_al_2013}
Solovchuk, M., Sheu, T., and Marc, T. (2013).
\newblock Simulation of nonlinear {Westervelt} equation for the investigation
  of acoustic streaming and nonlinear propagation effects.
\newblock {\em The Journal of the Acoustical Society of America}, 134:3931--42.

\bibitem[Tam and Webb, 1993]{Tam_and_Webb_1993}
Tam, C.~K. and Webb, J.~C. (1993).
\newblock Dispersion-relation-preserving finite difference schemes for
  computational acoustics.
\newblock {\em Journal of Computational Physics}, 107(2):262--281.

\bibitem[Treeby et~al., 2020]{Treeby_et_al_2020}
Treeby, B., Wise, E., Kuklis, F., Jaros, J., and Cox, B. (2020).
\newblock Nonlinear ultrasound simulation in an axisymmetric coordinate system
  using a k -space pseudospectral method.
\newblock {\em The Journal of the Acoustical Society of America},
  148:2288--2300.

\bibitem[Wang et~al., 2012]{Wang_et_al_2012}
Wang, K., Teoh, E., Jaros, J., and Treeby, B. (2012).
\newblock Modelling nonlinear ultrasound propagation in absorbing media using
  the k-wave toolbox: Experimental validation.
\newblock In {\em IEEE International Ultrasonics Symposium, IUS}, pages
  523--526.

\bibitem[Westervelt, 1963]{Westervelt_1963}
Westervelt, P.~J. (1963).
\newblock Parametric acoustic array.
\newblock {\em The Journal of the Acoustical Society of America},
  35(4):535--537.

\bibitem[Whitham, 1952]{Whitham_1952}
Whitham, G.~B. (1952).
\newblock The flow pattern of a supersonic projectile.
\newblock {\em Communications on Pure and Applied Mathematics}, 5(3):301--348.

\bibitem[Whitham, 1999]{Whitham_1999}
Whitham, G.~B. (1999).
\newblock {\em Linear and Nonlinear Waves}.
\newblock John Wiley \& Sons, Ltd.

\bibitem[Willatzen, 2001]{Willatzen_2001}
Willatzen, M. (2001).
\newblock Sound propagation in a moving fluid confined by cylindrical walls -
  {A} comparison between an exact analysis and the local-plane-wave
  approximation.
\newblock {\em Journal of Sound and Vibration}, 247:719--729.

\end{thebibliography}

\end{document}